\documentclass[12pt]{article}
\usepackage{epsfig, amssymb}
\usepackage{graphicx,psfrag} 
\usepackage{mathrsfs}
\usepackage{euscript}
\usepackage{bbold}
\usepackage{amsmath}
\usepackage[hang,flushmargin]{footmisc} 
\renewcommand{\baselinestretch}{1.4}
\DeclareMathAlphabet{\mathpzc}{OT1}{pzc}{m}{it}

\usepackage[ps2pdf,bookmarks=true,colorlinks,linkcolor=red,urlcolor=blue,citecolor=blue]{hyperref}
\usepackage[usenames]{color}

\newcommand{\cog}{}
\newcommand{\coo}{}

\def\tr{{\rm tr }}

\def\CA{{\cal A}}

\def\CF{{\cal F}}

\def\CM{{\cal M}}
\def\CN{{\cal N}}
\def\CO{{\cal O}}
\def\CP{{\cal P}}
\def\CQ{{\cal Q}}
\def\CK{{\cal K}}
\def\J{{\cal J}}

\def\II{\relax{I\kern-.10em I}}

\def\IZ{\relax{\rm Z\kern-.34em Z}}
\def\IB{\relax{\rm I\kern-.18em B}}
\def\IC{{\relax\hbox{$\inbar\kern-.3em{\rm C}$}}}
\def\ID{\relax{\rm I\kern-.18em D}}
\def\IE{\relax{\rm I\kern-.18em E}}
\def\IF{\relax{\rm I\kern-.18em F}}
\def\IG{\relax\hbox{$\inbar\kern-.3em{\rm G}$}}
\def\IGa{\relax\hbox{${\rm I}\kern-.18em\Gamma$}}
\def\IH{\relax{\rm I\kern-.18em H}}
\def\II{\relax{\rm I\kern-.18em I}}
\def\IK{\relax{\rm I\kern-.18em K}}
\def\IP{\relax{\rm I\kern-.18em P}}

\def\inbar{\,\vrule height1.5ex width.4pt depth0pt}

\def\IR{\relax{\rm I\kern-.18em R}}

\def\lp10{\ell_p^{10}}
\def\lp11{\ell_p^{11}}
\def\R11{R_{11}}

\def\frac#1#2{{#1 \over #2}}

\newcommand{\Qa}{Q}
\newcommand{\be}{\begin{equation}}
\newcommand{\ee}{\end{equation}}
\newcommand{\rs}{r_{\star}}
\newcommand*{\contains}{\rotatebox[origin=c]{-180}{$\in$}}%

\newdimen\tableauside\tableauside=1.0ex
\newdimen\tableaurule\tableaurule=0.4pt
\newdimen\tableaustep
\def\phantomhrule#1{\hbox{\vbox to0pt{\hrule height\tableaurule width#1\vss}}}
\def\phantomvrule#1{\vbox{\hbox to0pt{\vrule width\tableaurule height#1\hss}}}
\def\sqr{\vbox{%
  \phantomhrule\tableaustep
  \hbox{\phantomvrule\tableaustep\kern\tableaustep\phantomvrule\tableaustep}%
  \hbox{\vbox{\phantomhrule\tableauside}\kern-\tableaurule}}}
\def\squares#1{\hbox{\count0=#1\noindent\loop\sqr
  \advance\count0 by-1 \ifnum\count0>0\repeat}}
\def\tableau#1{\vcenter{\offinterlineskip
  \tableaustep=\tableauside\advance\tableaustep by-\tableaurule
  \kern\normallineskip\hbox
    {\kern\normallineskip\vbox
      {\gettableau#1 0 }%
     \kern\normallineskip\kern\tableaurule}%
  \kern\normallineskip\kern\tableaurule}}
\def\gettableau#1 {\ifnum#1=0\let\next=\null\else
  \squares{#1}\let\next=\gettableau\fi\next}

 \def\eqnn#1{\xdef #1{(\secsym\the\meqno)}\writedef{#1\leftbracket#1}%
 \global\advance\meqno by1\wrlabeL#1}
 \def\eqna#1{\xdef #1##1{\hbox{$(\secsym\the\meqno##1)$}}
 \writedef{#1\numbersign1\leftbracket#1{\numbersign1}}%
 \global\advance\meqno by1\wrlabeL{#1$\{\}$}}
 \def\eqn#1#2{\xdef #1{(\secsym\the\meqno)}\writedef{#1\leftbracket#1}%
 \global\advance\meqno by1$$#2\eqno#1\eqlabeL#1$$}

\global\newcount\itemno \global\itemno=0

\def\itemaut#1{\global\advance\itemno by1\noindent\item{\the\itemno.}#1}

\def\del{\partial}

\def\({\left(}
\def\){\right)}

\def\eg{{\it e.g.}}
\def\ie{{\it i.e.}}

\def\etal{{\it et.~al.}}
\newif{\ifeq}           
\eqtrue                 
\def\kX{M_{\text{BC}}}

\numberwithin{equation}{section}

\begin{document}

\begin{titlepage}

\begin{flushright}
MIT-CTP/4299, NSF-KITP-11-224
\end{flushright}
\vskip.2in
\begin{center}
{\huge String theory duals}\\
 {\huge of}
 \\
{\huge Lifshitz-Chern-Simons gauge theories}\\
\end{center}
\vskip.3in
\begin{center}
{\large Koushik Balasubramanian and John McGreevy}\\
Center for Theoretical Physics, MIT, Cambridge, MA 02139, USA\\
KITP, Santa Barbara, CA 93106, USA
\end{center}
\begin{center}
{\large Abstract}
\end{center}

{
We propose candidate gravity duals for a
class of non-Abelian $z=2$ Lifshitz Chern-Simons (LCS)
gauge theories studied by Mulligan, Kachru and Nayak. These are
nonrelativistic gauge theories in 2+1 dimensions in which
parity and time-reversal symmetries are explicitly broken by the
presence of a Chern-Simons term. 
We show that these field theories
 can be realized as deformations of DLCQ ${\cal N}=4$ super Yang-Mills theory. 
Using the holographic dictionary, we identify the bulk fields of type IIB supergravity that are dual
to these deformations.
The geometries describing the groundstates of the non-Abelian LCS
gauge theories realized here exhibit a mass gap.
}


\noindent
\vspace{3mm}

\end{titlepage}
\newpage
\renewcommand{\baselinestretch}{1.1}  

\renewcommand{\arraystretch}{1.5}

\section{Introduction}
\label{sec:intro}
Recently,  Mulligan \etal~\cite{Mike} studied an Abelian gauge theory in 2+1 dimensions with $z=2$ Lifshitz scaling symmetry, $x \to \lambda x, t \to \lambda^2 t$. Parity symmetry is explicitly broken by the presence of a Chern-Simons term in this theory. 
Unlike the Maxwell kinetic term in Maxwell-Chern Simons theory,
the Lifshitz-type kinetic term is marginal. Hence, the Chern-Simons term and the Lifshitz kinetic term compete in the infrared (IR). The usual Maxwell kinetic term is a relevant operator in this non-relativistic theory and  it tunes the system through a quantum phase transition between isotropic and anisotropic quantum Hall states. The critical point is reached by tuning the coupling to this operator to zero and is described by $z=2$ Lifshitz-Chern-Simons (LCS) theory. A non-Abelian extension of this theory (so far without Chern-Simons term) is currently under investigation \cite{MikeNA}. 

We would like to understand if this non-Abelian Lifshitz theory could be holographically related to a gravity theory in 
asymptotically $z=2$ Lifshitz spacetime \cite{KLM}. The string theory 
embeddings of Lifshitz solutions studied in \cite{BN,LifString,Lifstring2,Cassani:2011sv}, can be helpful in this 
regard\footnote{In this paper we deal with the $z=2$ case only.  \cite{Lifstring2} found Lifshitz solutions of massive type IIA and type IIB supergravity for a general dynamical exponent $z$. }. 
Some of these solutions can be described as deformations of DLCQ (discrete light-cone quantization) of ${\cal N} = 4$ super Yang-Mills (SYM) theory\footnote{The construction in \cite{BN} is a DLCQ (discrete light-cone quantization) of ${\cal N} = 4$ SYM theory, with a coupling that depends on the compact null direction. This non-trivial behavior of the coupling breaks the non-relativistic conformal symmetries of DLCQ ${\cal N} = 4$ theory to Lifshitz symmetries. Such a deformation of ${\cal N} = 4$ theory must result in a 2+1 dimensional non-Abelian Lifshitz theory with matter fields (in the adjoint representation). 
However, for our purposes, the non-trivial behavior of the coupling complicates the study of this effective 2+1 dimensional theory.}.
We will argue below (in section \ref{sec:qft}) that the solutions studied in \cite{LifString} are holographically dual to non-Abelian LCS theories with matter fields (which are organized into supermultiplets).
In order to make a connection with the non-Abelian LCS theory discussed in \cite{MikeNA}, these additional matter fields must be lifted.
 We will now discuss an approach for finding gravity duals of LCS gauge theories without additional matter. 

To obtain a 2+1 dimensional field theory, 
let us consider the theory on $N$ D3-branes with one longitudinal direction compactified into a circle,
whose coordinate we denote $x_3$, $x_3 \equiv x_3 + L_3$. 
Following \cite{Witten}, impose anti-periodic boundary conditions (APBCs) for the fermions and periodic boundary condition on the bosons. This boundary condition breaks supersymmetry and makes the fermions massive; the bosons then get mass through loop corrections. The masses of the bosons and fermions are of the order of inverse radius of the circle. 
For energies lower than the mass of the fermions
the theory is effectively 2+1 dimensional and described by pure Yang-Mills theory.  Type IIB supergravity in the $AdS$ soliton solution is holographically dual to the confining
groundstate of the theory described here \cite{Witten},
with the caveat that at large 't Hooft coupling the Kaluza-Klein modes cannot be parametrically decoupled.
We will see below that the Lorentz-violating system of interest in this paper allows an extra parameter.

Let us now deform this 2+1 dimensional theory by introducing a $\theta$ term 
that varies linearly along $x_3$, $\theta = k x_3/L_3$.  This deformation produces a Chern-Simons term in the effective  2+1 dimensional theory:
\be \int \theta~\text{tr} \left( F \wedge F \right) = \int d\theta \wedge \left( A \wedge d A + {2 \over 3} A \wedge A\wedge A \right) = k\int  \left( A \wedge d A + {2 \over 3} A \wedge A\wedge A \right) ~.\ee
In the above equation we have integrated by parts to get the first equality,
and neglected dependence on $x_3$ to get the second.
In the string theory dual description, this deformation corresponds to turning on $k$ units of 
RR-axion flux around the circle.
  We would like to know how the bulk geometry gets modified when this deformation is turned on.
If we assume that the axion flux is small, then its backreaction on the metric can be neglected. 
However, the circle shrinks to zero size at the tip of the soliton, and hence there must be a source for the axion flux at the tip of the soliton.
 This suggests that the axion flux is sourced by 
 D7 branes at the tip of the soliton which in principle resolves the conical singularity induced by the axion flux. This singularity is resolved when the number of D7 branes equals the axion flux. The presence of D7 branes makes the IR behavior different from that of the ``undeformed'' $AdS$ soliton background. A related discussion appears in \cite{RyuTakayanagiFQH} as a holographic model of fractional quantum Hall systems.
Note that this construction is similar to a holographic realization of ${\cal N}=1$ super Yang-Mills-Chern-Simons 
theory found in \cite{MN}.

 We can now give an alternate interpretation{\footnote{\cog{This intepretation is based on the following facts:
D7 branes at the tip will introduce non-trivial monodromy for the RR-axion which extends into the UV. So the information from the D7 branes at the tip should be considered as UV data. This monodromy results in a Chern-Simons term in the 2+1 dimensional gauge theory. 
On the other hand, such a CS term arises from integrating out massive fermions transforming in the fundamental representation.}}}  of the low energy effective theory described earlier.
The identity of the flat-space brane system whose near-horizon limit we want is not clear,
but we can infer a few ingredients.
The addition of D7 branes corresponds to addition of matter multiplets in the boundary theory;
these 
multiplets transform in the fundamental representation of the $SU(N)$ gauge theory living on the D3 branes. The strings stretching between the D3 and D7 branes are massive. The 2+1 dimensional effective theory we get by integrating out these massive modes is a 2+1 dimensional YMCS theory (see \eg~\cite{Bergman}, \cite{ABJM}).     
 
 In the large $k$ ($k\gg1$) limit, we can utilize ideas of geometric transition to replace the D7 branes at the tip of the soliton by axion flux in the background of a ``deformed soliton". Such a ``deformed soliton" must be regular everywhere with the $x_3$ circle being topologically non-trivial.
 By analogy with the conifold and other examples, we might expect that the $S^5$ should become trivial in the IR in the deformed geometry. 
Other possibilities are that the non-compact part of the metric 
 could be multiplied by a warp factor that has a minimum in the IR,
 or that the dilaton profile could become singular in the IR. It appears that such a ``deformed soliton'' (if it exists) is dual to ${\cal N} =0$ YMCS theory.

Now, let us consider a situation where the circle with axion flux is non-trivially fibered over 
the space, time and radial directions (which are denoted $\vec x, t, r$ respectively, below):
$$ ds^2 = e^{2 \sigma} \( dx_3 + \CA \)^2 + ds_4^2 $$
where $ds_4^2$ denotes the line element in the $\vec x, t, r$ directions.
Such solutions are in general dual to 2+1 dimensional non-relativistic Chern-Simons theories. 
A new possibility arises here 
where $ g_{33} = e^{2 \sigma} $ vanishes in the IR,
but the fiber does not degenerate,
in the sense that $ e^{2\sigma} \CA dx_3 $ is nonzero.
In this case, the $x_3$-circle can carry an axion flux,
but (if the geometry is non-singular) there are no additional sources (such as D7-branes) for the axion field. 
 
\begin{figure}[htbp]
   \centering
   {\includegraphics[scale=0.6]{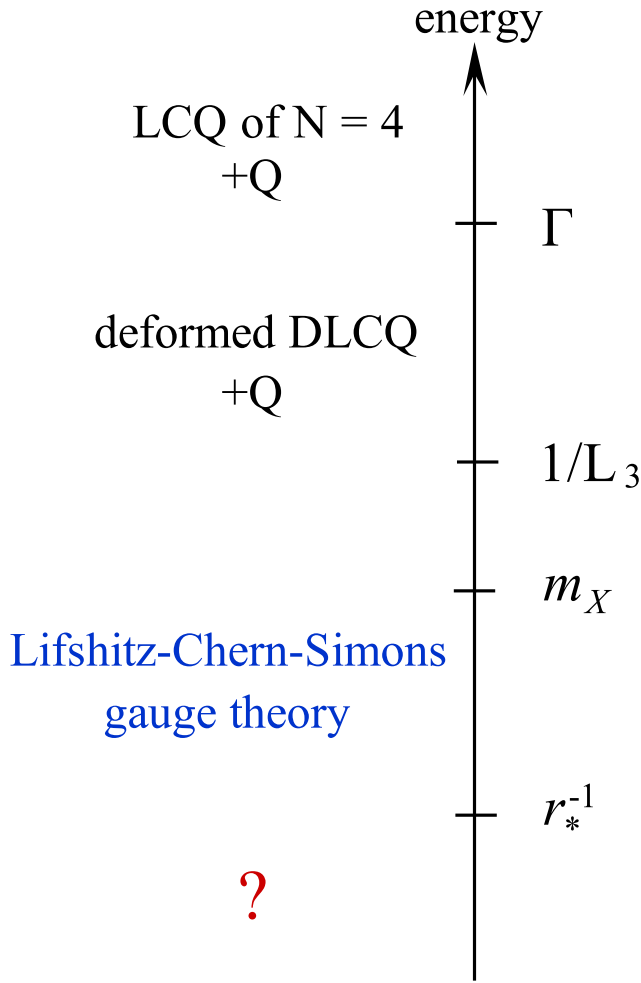} 
   }
   \caption{\label{fig:scheme}
The hierarchy of energy scales considered in this paper.
The highest scale is $\Gamma$, which is the coefficient in 
the field theory action of certain protected-dimension operators which would be irrelevant 
as perturbations of the $\CN=4$ theory with the ordinary $z=1$ scaling.
In the bulk solution $\Gamma$ (defined in \eqref{eq:gammadef}) appears via ordinary non-normalizable falloffs of the bulk metric.
The next scale from the UV is the inverse-radius of the circle;
at this scale the APBCs break supersymmetry.
The scale $m_X$ is the one we control the least.
It is determined by a mass scale
(called $M_{\text{BC}}$ below)
encoded in the metric boundary conditions;
in the field theory, this is the coefficient of
a certain operator whose dimension is not protected by supersymmetry.
Below $m_X$, the physics is described by 
the Lifshitz-Chern-Simons gauge theory of interest.
The scale $r_\star^{-1}$ is referred to in the paper as ``the IR scale".
The question mark is the subject of \S\ref{sec:repair}.
   }
\end{figure}

 In this paper, we will study non-relativistic solutions with 
 properties of the ``deformed soliton" described above. These solutions 
 arise via a null deformation{\footnote{A null deformation is defined to be a deformation 
 of a supergravity solution that preserves a null Killing vector.}} of type IIB on $AdS_5 \times S^5$ (with one of the lightcone coordinates compactified) and hence the dual field theory is a deformation of DLCQ of ${\cal N}=4$ SYM theory. 
In contrast to the story in the AdS soliton, the IR scale (holographically, this is the point at which the $x_3$-circle 
shrinks) in this solution 
is not determined by the compactifcation size of the shrinking circle.
 Rather, there is a second mass scale in the problem, in addition to the inverse radius of the compact circle. 
 A precedent for this situation is the
  confining solution studied in \cite{Gubser}, where the confinement scale is determined by boundary conditions on the dilaton. The boundary condition on the dilaton introduces an an additional mass scale, for which Gubser \cite{Gubser} provided a field theory interpretation. We will use this idea to argue that the deformed $\CN=4$ SYM theory dual to our solution is described by non-Abelian $z=2$ LCS theory at low energies (below
  the KK scale $1/L_3$, and above the dynamical scale -- see Fig.~\ref{fig:scheme}). The low energy effective theory in this regime 
  inherits the $z=2$ scaling symmetry of DLCQ ${\cal N}=4$ SYM theory, but it is not Galilean invariant. 
  
The usual problem of DLCQ is the lightcone zeromodes \cite{Hellerman:1997yu}.
They generally produce an infinitely-strongly-coupled static sector
of the theory which must be solved first.
Remarkably, here these zeromodes conspire to comprise exactly the 
auxiliary fields in the first-order description of the Lifshitz gauge theory \cite{Mike}.
These solutions seem to be dual to a pure glue
theory in a wide range of energy scales.
The freedom due to broken Lorentz invariance
allows us to decouple the IR scale from the Kaluza-Klein scale.

The rest of the paper is organized as follows. 
In section \ref{sec:nulldef}, we will present a solution with RR axion flux which has asymptotic $z=2$ Lifshitz scaling symmetry. In section \ref{sec:qft}, we will argue that the solution in section 
\ref{sec:nulldef} 
is dual (in a window of energies)
to large-$N$ non-Abelian LCS theory. 
In particular, we will show that LCS theories can be realized as deformations of DLCQ ${\cal N}=4$ SYM theory. 
The solution of \S\ref{sec:nulldef} is not geodesically complete \cite{Myers, Copsey},
for geodesics with sufficiently large momentum around the circle $p^3$.
It is nevertheless useful for studying the physics of modes with $p^3=0$;
in section \ref{sec:gap}, we study the dependence of the spectrum of glueballs with $p^3=0$
on the Chern-Simons level as a consistency check.
In the final section \ref{sec:repair}, we provide two resolutions of the problems raised by 
\cite{Myers, Copsey}.  One (in \ref{sec:rep1}) is a realization 
of dilaton-driven confinement that follows upon perturbing 
the previously-described system by a certain (dangerously-irrelevant) operator.
The other (in \ref{sec:rep2}) looks like a Higgs vacuum of the theory,
and is a candidate for the true groundstate.
Several appendices sequester background information and technical details.
In appendix \ref{sec:dilatondriven}, we will briefly review
dilaton-driven confinement, primarily based on \cite{Gubser}.
The remaining appendices give
a family of confining solutions (\ref{app:solutions}), 
an analysis of the supersymmetry which would be preserved 
if we used periodic boundary conditions (\ref{app:susy}), 
and a detailed analysis of the UV boundary data (\ref{app:counterterms}).

\section{Null deformations of $AdS_5 \times S^5$}
  \label{sec:nulldef}
  
 In this section we will study solutions of type IIB supergravity that can be obtained as null deformations of $AdS$. 
 In order to study such solutions it is convenient to parametrize $AdS$ by light cone coordinates.
Compactifying one of the null directions is a simple example of a null deformation. This example does not alter the form of the metric or other supergravity fields but  changes the boundary conditions. A compact null direction can be obtained from a compactified spatial direction by an infinite boost along the compact direction. In general, a null deformation modifies the supergravity fields in such a way that one of the null directions becomes spacelike.  
 
In this paper we will study a particular class of such deformations. The following is a solution of the type IIB equations of motion (see \eqref{TypeIIBEOM} for conventions):
\be ds^{2}_{\text{Lif}} = L^{2}\({2 dx_{3} dt + d\vec x^{2} + dr^{2} \over r^{2}} +f(r) dx_{3}^{2}  + ds^{2}_{S^{5}}\), \quad f(r) =  {\Qa^{2} e^{2\Phi_{0}}  \over 4L_{3}^{2} } - \left({r^{2} \over r_0^{4}}\right) \label{Lifcut} \ee
$$F_{5} =2L^{4}(1+ \star)\Omega_{5} , \quad C_{0} = {\Qa x_{3} \over L_{3}}, \quad \Phi = \Phi_{0}  $$
In the above solution, $x_{3}$ is a compact direction $(x_{3} \equiv x_{3} + L_{3})$. Translation symmetry along $x_3$ is broken by the RR-axion profile $\chi(x_3)$. Similar solutions have been studied in the context of string embeddings of Lifshitz spacetime. 
In fact, the solution described above has asymptotic Lifshitz symmetries.{\footnote {Specifically, 
the following scaling symmetry is an asymptotic isometry of (\ref{Lifcut})
 $$t \rightarrow \lambda^{2}t,\quad \vec{x} \rightarrow \vec{x}, \quad r\rightarrow \lambda r, \quad x_{3} \rightarrow x_{3}.$$
This is also a symmetry of the geometry with non-compact $x_3$. }}
This geometry approaches $\text{Lif}_{z=2}^{d=2}$ geometry
in the UV ($r \rightarrow 0$).
$x_{3}$ is a compact direction and hence cannot scale, as scaling will change the compactification radius. This solution is not invariant under time reversal symmetry and parity. The non-invariance of the solution under parity 
cannot be seen in the geometry, but it can be inferred from the non-trivial profile for the RR-axion, which is a pseudo-scalar.

It is not possible to take $Q \rightarrow 0$ in the above solution, while fixing $r_0$
and maintaining regularity.  Further, when $Q = 0$ and $r_{0} \rightarrow \infty$, translation symmetry along $x_3$ is restored. Hence, we have asymptotic Lifshitz symmetries only when $Q\neq 0$.

At this point we pause to anticipate a crucial point about the role of the 
scale $r_0$ in the solution \eqref{Lifcut}.  
While it determines the (IR) location where the $x_3$ circle shrinks,
we demonstrate below (in appendix~\ref{app:counterterms}) that $r_0$ is a non-fluctuating quantity
that should be considered a part of the short-distance definition of the theory.
In particular, it is introduced as the coefficient of a 
boundary counterterm at the UV cutoff surface, whose addition is required
to establish a well-posed variational problem of which \eqref{Lifcut} is a solution.
This will be understood on the QFT side as the extra UV data required to make sense 
of a deformation of a CFT by an irrelevant operator.


The IR behavior of the geometry \eqref{Lifcut} is a bit subtle.  
The $x_3$ circle shrinks
in the IR when $f(r) =0$, {\it i.e.}, when
\be 
\label{eq:rstar}
r= r_{\star}={r_{0}^{2} } \left({\Qa e^{\Phi_{0}} \over 2 L_{3}}\right) .\ee 
There is no conical singularity at 
this locus.
All curvature invariants of the metric in (\ref{Lifcut})
are finite, as a consequence of $g^{33}$ being zero. 
The geometry is free of curvature and conical singularities. 

After the first version of this paper appeared on the arXiv, 
 it was pointed out \cite{Myers, Copsey} that the metric in (\ref{Lifcut}) is geodesically incomplete if the radial coordinate 
 is restricted to lie between $0$ and $r_{\star}$. In particular, it was shown that certain geodesics carrying non-zero momentum along $x_{3}$ do not lie entirely in the region $ r< r_{\star}$. A straightforward way of extending these geodesics past $r_{\star}$ leads to closed timelike curves (CTCs) in this region. 
Clearly, it is important to understand the implications of this hidden singularity on the dual field theory.  
We discuss this issue and its resolutions in \S\ref{sec:repair}. 
Until then, we focus on 
the problem of identifying 
the field theory
at energies above $r_\star^{-1}$, where the solution (\ref{Lifcut}) is less problematic.

Geodesics carrying zero momentum along $x_{3}$ direction do not cross $r_{\star}$ and hence the physics of the zero modes may be insensitive to this region.
The fact that modes with $p^3=0$ do {\it not} penetrate past $r=r_\star$
suggests that for the purposes of the 2+1-dimensional physics of interest to us in this paper 
we may terminate the geometry at $r=r_\star$.
This in turn suggests that the dual field theory (at least the spectrum of operators with $p_3=0$) is gapped. 
The energy scale associated with the gap
$r_{\star}^{-1}$ 
(we will compute the energy gap for these modes in more detail below)
is not determined by specifying $L_{3}$ alone. 
This is not unprecedented.
In appendix \ref{sec:dilatondriven}, we review confining solutions where the confinement scale is not determined by the radius of the shrinking circle. By analyzing the boundary counterterms in appendix~\ref{app:counterterms}, we conclude that $r_{0}$ is determined by the boundary conditions on the metric (or vielbeins). The discussion in appendix \ref{sec:dilatondriven}
indicates that the parameter $r_{0}$ produces to a mass deformation $m_X$ in the dual field theory.    
 
 
 Before we proceed to analyze the dual field theory, let us make some more comments about the solution.
T-duality along $x_3$ produces a solution of massive type IIA supergravity. 
This equivalent description clarifies some aspects of the physics
(though questions regarding the regularity of the solution are obscured).
For instance, it is easier to check that the T-dualized solution has asymptotic Lifshitz symmetries. We can also see that $r_0$ is determined by boundary conditions on the dilaton (of massive type IIA). The T-dualized solution has a non-trivial flux associated with the NS-NS $B$- field. Further, this $B$ field has a mass (determined by axion flux in the type IIB solution). This $B$ field is related to the $g^{t3}$ component of the IIB metric by T-duality. This suggests that the fluctuations of $g^{t3}$ (in the presence of an axion flux) satisfies a massive wave equation. Further, we will see that this field is dual to a dimension 6 operator of ${\cal N} = 4 $ SYM theory. The theory obtained by deforming ${\cal N} =4$ SYM theory by this dimension 6 operator is very similar to the non-commutative SYM theories studied in \cite{PFT}. The $S^{5}$ factor remains unaffected by these deformations. These observations will be helpful in analyzing the dual field theory.
 
 Let us now try to guess what the dual field theory (at low energies) could look like. At this point, we will not try to relate the parameters of the solution to the parameters of the field theory. We know that the solution has asymptotic Lifshitz symmetries. The presence of a shrinking circle suggests that the fermions must satisfy anti-periodic boundary conditions making them (and the scalars) massive. This suggests that the low energy theory is described by a 
 Lifshitz-symmetric pure gauge theory. We can also infer (from the profile for RR-axion) that the dual field theory should contain a Chern-Simons term. So, it appears that the dual field theory is a non-Abelian version of Lifshitz Chern Simons theory at low energy. In the next section, we will present detailed arguments supporting this claim. 
 
\section{Identification of the dual field theory}
\label{sec:qft}
 
In this section we will argue that the field theory dual to (\ref{Lifcut}) is described by a non-Abelian LCS theory in a range of energies 
(above the IR scale, below the KK scale).  
Here is the strategy: 
first we study the UV asymptotics and conclude that the QFT is a deformation
of the DLCQ of $\CN=4$ SYM.  
We organize the possible deformations to the QFT action by their scaling dimensions appropriate
to the DLCQ theory, and
identify the bulk fields to which they are dual following the extensive literature on AdS/CFT for 
$\CN=4$ SYM.  
This analysis can be done in the theory with $x_3$ noncompact (keeping the DLCQ scaling law in mind).
We find a gauge theory coupled to fermions and scalars.
Then we compactify $x_3$ 
to lift the fermions, and deform the boundary conditions on various supergravity modes
to lift the scalars.
As with any discussion of DLCQ, 
a tricky step in this analysis is the treatment
of the zeromodes around the $x_3$ direction.  
By APBCs, the fermions have no such zeromodes.
The scalar zeromodes are lifted by the mass deformation $m_X$.
We are left with the zeromodes of the gauge field.
We show that these organize themselves into 
a first-order description of the Lifshitz-Chern-Simons gauge theory.

To begin, let us note that the $x_{3}$-direction becomes null as we approach the boundary.  The dual field theory lives on the conformal boundary which is $ds^{2}_{bdy} = 2 dx_{3} dt + d\vec{x}^{2}$.{\footnote{ The notion of a conformal boundary need not be well-defined for non-relativistic backgrounds (for instance, Schr\"odinger spacetime \cite{Sch} is not conformally compact). Recently, a notion of anisotropic conformal infinity was introduced in \cite{Horava, RossHolRen, Boer, McNees} for non-relativistic spacetimes that are not conformally compact in the conventional sense. We would like to point out that the metric in (\ref{Lifcut}) is conformally compact unlike the Schr\"odinger spacetime. The conformal boundary is given by $$ ds^{2}_{bdy} = \mathop{\lim}_{r \rightarrow 0} {r^{2} \over L^{2}} ds^{2}_\text{Lif} =  2 dx_{3} dt + d\vec{x}^{2}.$$ }} 
This is just Minkowski space in lightcone coordinates.
Quantization of a field theory with $x_3$ compact, and $t$ treated as the time variable
is DLCQ.  A scale transformation under which $\vec x \to \lambda \vec x$ requires $ t \to \lambda^2 t$
to preserve the metric,
and hence $z=2$.

Since the solution (\ref{Lifcut}) differs from \eqref{Lif0} by non-normalizable field variations, 
the field theory dual is a deformation of the DLCQ of ${\cal N }=4$ SYM theory.
Operators that are irrelevant to the relativistic ${\cal N} =4$ theory can be marginal or relevant in the deformed DLCQ theory.
In order to study the dual field theory we must include irrelevant (with respect to $z=1$ scaling) deformations of ${\cal N} =4$ theory. We can ignore deformations that are irrelevant with respect to $z=2$ scaling symmetry as well;
this means operators with dimension greater than $8$ according the $z=1$ counting.

In light-cone YM theory, the equation of motion involving $F^{ti} \equiv F_{3i}$ is a constraint equation (Gauss' law). Hence, in the 2+1 dimensional non-relativistic theory (with $z=2$), $E^i \equiv F^{ti}$ appears as an auxiliary field with mass dimension $[E^{i}] =1$. 
In the non-relativistic theory marginal operators have mass dimension 4. This suggests that we must include terms of the form $\tr\(F_{3i}F_{3j}F_{3i}F_{3j}\)$ in the 3+1 dimensional theory. 
 
Now, let us consider the case where $r_{0} \rightarrow \infty$ and $x_{3}$ is non-compact. In this case, the solution preserves ${\cal N} =1$ supersymmetry (please see appendix~\ref{app:susy}). Hence, this solution is dual to a deformation of ${\cal N}=4$ SYM theory that preserves ${\cal N} =1$ supersymmetry\footnote{The $\CN=4$ theory in the presence of a linear axion profile,
and in particular the preservation of supersymmetry, have been studied recently in \cite{Gaiotto:2008sd}.}. The form of the metric suggests that this deformation breaks lightcone symmetry but preserves spatial rotation symmetry.  
The axion profile means that the deformation breaks parity.
The deformation preserves the $SO(6)$ invariance of the undeformed theory.
We will now identify the operators responsible for this deformation by studying the equations governing linear fluctuations of the supergravity fields. 

When $r_{0} \rightarrow \infty$, the line element (after reducing over the sphere) in (\ref{Lifcut}) can be written as follows {\cog{(we will set the $AdS$ radius $L=1$ from now on)}}
\be \label{Lif0}
ds^2_5 = {2 dx_{3} dt + d\vec x^{2} + dr^{2} \over r^{2}} +{Q^2 e^{2\Phi_{0}} \over 4L_3^2} dx_{3}^{2}~. \ee
Note that when $x_3$ is non-compact, $L_3$ is a length scale that makes the metric non-dimensional,
but it can be absorbed in a rescaling of the $x_3$ coordinate.
The asymptotic solution \eqref{Lif0} 
preserves four supercharges and 
is one studied in \cite{LifString}.  
According to the previous discussion, we see that it is dual to a supersymmetric Lifshitz-Chern-Simons theory.
The supergravity solution \eqref{Lif0} implies that the extra matter which allows for supersymmetry 
produces a conformal fixed point.  
{\coo{This holographic prediction merits further study.}}
In the following, we deform this theory by relevant operators
which lift the matter fields. 

 It is convenient to work with the vielbein formalism (see \cite{MTaylor}, \cite{Ross} for a discussion on the utility of this formalism in non-relativistic holographic renormalization).
In terms of vielbeins, the five dimensional metric takes the following form{\footnote{ We will use $\mu, \nu$ denote the spacetime indices $\{t,x_{1},x_{2},r\}$$~\mbox{and}  ~ a,b$ to denote the vielbein indices $\{0,1,2,4\} $ throughout this paper. Note that we will not use $3$ to denote any vielbein index as this denotes the label of the compact direction. We will use the letter $y$ to denote the fifth vielbein index.}}
  \begin{equation} ds^{2} = g_{\mu \nu} dx^{\mu} dx^{\nu} + {Q^2 e^{2\Phi_{0}} \over 4L_3^2} \left(dx^3 +  {4L_3^2 \over Q^2 e^{2\Phi_{0}} r^2}dt\right)^{2}  = \eta_{ab} e^a e^b + e^y e^y \end{equation} 
where 
\be 
e^4 =  dr/r,~
e^0 = {2 L_3 } {dt \over Q e^{\Phi_{0}}r^2},~ e^y = {Q e^{\Phi_{0}} \over 2 L_3} \left(dx_3 +  {4L_3^2 \over Q^2 e^{2\Phi_{0}} r^2}dt\right)
,~e^i = {dx^i\over r}~.\ee
Let us also define
$dx^\mu = \tilde{e}^{\mu}_a e^a $ and $dx^3 = \tilde{e}^3_{0} e^{0} + 
\tilde{e}^3_{y} e^{y} $. 
 {\footnote{Note that when we reduce along $x_{3}$, $e^{y}_{\mu}$ shows up as a vector field and $e^{y}_{3}$ as a scalar field in the lower dimensional theory. The non-trivial profiles for these fields are responsible for breaking Lorentz invariance in the lower dimensional theory. Some details about this reduction can be found in appendix~\ref{app:counterterms}.}} The operators we are interested in are dual to $e^{y}_{t}$ and $e^{y}_{3}$. We will now determine the dimensions of these operators by studying the equation that governs linear fluctuations of $e^{y}_{t}$ and $e^{y}_{3}$.  
Let us define $A' = \delta e^y_t dt$ and $\sigma  = \delta e^{y}_{3}$. The equations of motion for $\delta e^y_t$ and $\delta e^{y}_{3}$ can be written as{\footnote{The following relations were used to derive (\ref{fluct}): $$\del_y{\varphi} = \left(\del_3  + \tilde{e}^{\mu}_y 
\del_\mu \right) \varphi \quad \mbox{and} \quad \del_a{\varphi} = \left(\tilde{e}^3_a\del_3  + 
\tilde{e}^{\mu}_a \del_\mu \right) \varphi.$$ Note: In the definition of $\star_{4}$, $a_{i} \neq y$.}}
\be d\star_4 dA'  = m^2\star_4 A' \quad,\quad d\star_{4} d \sigma =0 \label{fluct}\ee
where $m^2 L^{2}= 16$ and $\star_4 (e^{a_{1}}\wedge \dots \wedge e^{a_{k}}) = e^{a_{k+1}}\wedge \dots \wedge e^{a_{4}}$. 
Note that the above equations are true even when $x^3$ is non-compact. 
We can see that $A'= r^{\Delta} A'_{0}$ is a solution of the above equation if $(\Delta-2)^2 = m^2$ or $\Delta_{\pm} = 2 \pm 4$. Hence, the dimension of the operator that is dual to this mode is 
$ \Delta_{\cal O} = 6$. 
The equation of motion for $\sigma$ suggests that $e^{y}_{3}$ is dual to a dimension 4 operator. 
Note that $A'$ is massive due to the presence of axion flux. 
A similar observation was made in \cite{SRDas} where the fluctuations of NS-NS field becomes massive due to the presence of five form flux. They showed that these fluctuations correspond to a dimension 6 operator in ${\cal N} = 4$ SYM theory.{\footnote{The authors identified this operator (anti-symmetric part) by expanding DBI and WZ action (for $N$ D3 branes). Their analysis was restricted to the case where $SO(6)$ invariance is not broken.}}   
The operator dual to the 2-form field is antisymmetric in Lorentz indices. In \cite{Ferrara}, it was shown that this dimension 6 operator lives in a short supermultiplet with $\tr\(W_{\alpha}W^\alpha \bar{W}_{\dot{\alpha}}\)$, where $W_{\alpha}$ denotes 10D $\CN =1$ superfield strength.{\footnote{\cog Here the undotted index denotes left-handed spinor and the dotted index  denotes right-handed spinor.}}

The operator dual to $e^{y}_{t}$ belongs to the same short multiplet and it can be written in terms of 10D $\CN =1$ superfields as follows{\footnote{We can write this in terms of 4D fields after reducing on a $T^{6}$.}}
$$ \tilde{{\cal O}}_6 = \int d^4 \theta E^{{\dot \alpha}}\mbox{tr} \left(W_\alpha W^\alpha \bar{W}_{{\dot \alpha}}\right) + h.c$$
where the boundary value of $e^{y}_{t} r^{-\Delta_{-}}$ has been promoted to a superfield $E^{{\dot \alpha}}$. When this operator is written in terms of the component fields, it must take the form ${\CO}^{t}_{\mu}\tilde{\underbar {\it e}}^{\mu}_{y} + {\CO}^{t}_{3}\tilde{\underbar {\it e}}^{3}_{y}$, where $\tilde{\underbar {\it e}}^{\mu}_{y}$ and $\tilde{\underbar {\it e}}^{3}_{y}$
denote the coefficients of non-normalizable fall-offs of $\tilde{e}^{\mu}_{y}$ and $\tilde{e}^{3}_{y}$ respectively. Further, we know that $\tilde{e}^{\mu}_{y} = 0, g^{bdy}_{33} = g^{bdy}_{tt} =0$. Using these facts we can see that the operator that is dual to $e^{y}_{t}$ is of the form ${\CO}_{x_{3}x_{3}}$.
Following \cite{SRDas,Ferrara}, we can write down this dimension 6 operator that is dual to $e^y_{t}$
$$ {\cal O}_{6} = i\mbox{tr} \left( [F_{3k},F_{l3}]F^{kl} + F_{3k}\partial_3 X^I \partial^k X^I \right) + \mbox{terms involving fermions.}$$
Similarly, we can show that the operator dual to $e^{y}_{3}$ is 
$$ {\cal O}_{4} = T_{3t} = \tr \left( F_{3i}F_{ti} - {1\over 4} F^{2} \) + \mbox{terms involving fermions and scalars.} $$
This operator belongs to the short multiplet $\tr\(W_{\alpha}\bar{W}_{\dot{\alpha}}\)$. Note that  $\tr \left( F_{3i}F_{ti} \)$ appears as a kinetic term in the lower dimensional non-relativistic theory. Now, the only other $SO(6)$ invariant operator with dimension $ \Delta \le 8$ is the operator dual to the volume form of $S^{5}$. This operator has dimension 8 and it has been identified in \cite{dim8, Ferrara} to be
$$\tilde{\cal O}_{8}=\mbox{tr} \left(F_{IJ}F_{KJ}F_{IL}F_{KL}+{1\over 2}F_{IJ}F_{KJ}F_{KL}F_{IL} -{1\over 4}(F^{2})^{2} \right) $$ $$ \qquad \quad + \quad \mbox{terms involving fermions and scalars.}$$
This operator lies in $\tr\(W^{2}_{\alpha}\bar{W}^{2}_{\dot{\alpha}}\)$ and hence its dimension is protected. It was conjectured in \cite{dim8} that moving away from the near-horizon geometry of D3 branes corresponds to deforming the ${\cal N} =4$ theory by the dimension 8 operator $\tilde{\cal O}_{8}$. This operator is irrelevant with respect to $z=1$ scaling and its effects disappear from the dual field theory when we take the strict near-horizon limit.    

Equation (\ref{Lif0}) describes the IR geometry of plane-wave-deformed D3 brane geometry \cite{Hartong}. 
Including deviations away from the IR region 
of a D3 brane geometry 
(which would ultimately glue it to the asymptotically flat $\IR^{9,1}$)
should correspond to adding the dimension 8 operator $\tilde{\CO}_{8}$ \cite{dim8}. When $x_{3}$ is compact and therefore does not scale, 
terms of the form $\CO_8 \equiv \mbox{tr} \left(F_{3j}F_{3j}F_{3i}F_{3i}\right)$ 
are not suppressed in the strict low-energy limit. 
Such terms
in $\tilde{\CO}_{8}$ cannot be ignored in the low energy effective theory 
that is dual to (\ref{Lif0}), with compact $x_3$.


When $x_{3}$ is non-compact, the theory that is dual to (\ref{Lif0}) is 
therefore described by the following action {\footnote{\cog{This theory describes the low energy limit of the world volume theory of  $N$ D3 branes deformed by a linear $\theta-$term. In the low-energy theory the ${\CO}_{6}$ is present as irrelevant deformations (w.r.t $z=1$ scaling) 
In fact this operator will be generated through loop corrections due to the presence of the non-constant $\theta-$term. Further, there is no symmetry preventing this operator from appearing in the low energy theory.
 }}}
\begin{equation} S_{1} = S_{{\cal N}=4}(\theta = Q x^{3}/L_{3}) + \int dt d^{2}x dx_{3} \left[\kappa_6 {\cal O}_6 + \kappa_8 {\cal O}_8 + \dots \label{ncom} \right]\end{equation}
where $  {\cal O}_8 = \mbox{tr}\left([F_{3i},F_{3j}]^{2}\right) + \mbox{terms involving scalars and fermions}$, and $\theta$ is the theta-angle of $\CN = 4$ theory. We will not worry about the operators denoted by $...$ as these are irrelevant with respect to both $z=1$ and $z=2$ scaling.
Note that $\kappa_{6}$ has mass dimension $-2$. 
The non-normalizable fall-off of $e^y_t$ suggests that the coupling is proportional to $ L_3^2/Q^2$.  
Before we compactify $x_{3}$ let us define :
 $$F^{ti} = F_{3i} = \sqrt{\kappa \over \kappa_{6}}  E_{i}, \quad F^{3t} = F_{t3} = E_{3}, \quad {g'}^{2}_{1} = g_{1} L_{3} = g^{2} \sqrt{\kappa_{6} \over \kappa}, \quad \lambda_{1} = \kappa_{8} {\kappa^{2} \over \kappa_{6}^{2}}, \quad g_{2} = g_{3} =g $$
 The action $S_{1}$ when written in terms of the new variables reads as follows
$$S_{3+1} \equiv  \int dt d^{2}x dx_{3} \Bigg[{1\over 2{g'}_{1}^{2}} \mbox{tr} \left( E_{i}D_{t}A_{i} + A_{t}D_{i}E_{i}\right) + {1\over 4g_{2}^{2}}\mbox{tr}\left(F_{ij}F^{ij}\) + {1\over 2g_{3}^{2}}\mbox{tr}\left(E_{3}^{2}\right)  +  $$ \begin{equation} \lambda_{1} \mbox{tr}\left([E_{i},E_{j}]^{2}\right)+ i \kappa \mbox{tr}[E_{i},E_{j}]F^{ij}   + \mbox{terms involving scalars and fermions} \Bigg] \label{Th1}\end{equation} $$ + {Q\over L_{3}} \int dx_{3} \wedge \mbox{tr}\(A \wedge F\) $$
We can see that this action resembles a gauge theory action written in first order formalism. Also, note that ${g'}_{1}^{2}$ has mass dimension -1. 

At last we consider the theory with compact $x_3$.  
The field theory dual of (\ref{Lifcut}) is a deformation of (\ref{Th1}). 
Compactifying $x_{3}$ with anti-periodic boundary conditions on the fermions makes them massive.  
Kaluza-Klein reduction 
along $x_{3}$
of the last term in \eqref{Th1}
 induces a Chern-Simons term in the effective 2+1 dimensional theory. 
We can absorb the overall factor of $L_{3}$ by rescaling $t$. 

The IR scale in the geometry $r_{0}$ is determined by non-trivial boundary behavior of $e^{y}_{3}$ and $e^{y}_{t}$ (see appendix~\ref{app:counterterms}).
The discussion in appendix \ref{sec:dilatondriven} suggests that the non-trivial boundary conditions on $e^{y}_{3,t}$ are induced by some excited string state. 
The end result is a mass term $m_X$ for the scalars, presumably by operator mixing.
The heuristic calculation in appendix \ref{sec:dilatondriven} suggests 
that $\kX \sim {m_{X} \over \Delta_{K}(\lambda)}$ which gives $\kX \ll m_{X}$ 
at large 't Hooft coupling.  
The placement of $m_X$ in Fig.~\ref{fig:scheme} is based on this estimate;
unfortunately we do not know the precise relationship between $m_X$ and the other scales in the problem.

We are interested in the low energy effective description for energies less than $m_{X}, L_{3}^{-1}$.  
We see that the modes with non-zero Kaluza-Klein momentum are massive (with mass $\ge L_{3}^{-1}$),
and can be integrated out. Hence, the low energy dynamics is described by the dynamics of the 
modes with no dependence on $x_3$. 
Because of the APBCs, there are no modes of the fermion fields with this property.
We will now show that the scalar zero modes can also be integrated out 
with impunity for energies less than $m_{X}$. To see this, let us study the behavior of the scalar zero mode propagator when the proper radius of $x^{3}$ in the boundary theory is $L_{prop} \sim \epsilon_{3} L_{3}$. The propagator of the DLCQ theory is obtained by taking $\epsilon_{3} \rightarrow 0$. When $\epsilon_{3}$ is small but non-zero, the zero modes are dynamical and the momentum space propagator of the zero mode is given by
$$ D(\omega, \vec{k}) \sim \(\epsilon_{3}^{2}\omega^{2} -\vec{k}^{2} -m^{2}_{X}\)^{-1}$$  
Now, for $\omega < m_{X}$, we can integrate out the zero mode without introducing divergences in the Feynman graphs containing zero mode propagators. When $m_{X}$ is zero, the zero modes do introduce divergences at $\vec{k} =0$; in other words, the scalar zero mode gets strongly coupled with other zero modes and non-zero modes. Note that the zero modes are problematic when we try to quantize the 3+1 dimensional theory using DLCQ
\cite{Hellerman:1997yu}
and the symptom is this divergence. 
This is because, to quantize the 3+1 dimensional theory, we need to quantize all non-zero modes. However, only the modes with mass less than $m_{X}/\epsilon_{3}$ can be decoupled from the zero mode. Modes with mass greater than $m_{X}/\epsilon_{3}$ get strongly coupled with the zero mode. However in our case, we are interested in finding the effective description for energies less than $m_{X}$. Hence, the scalar zero modes can be decoupled without introducing divergences.  

So the only degrees of freedom for energies less than $m_{X}$ are the zero modes of the gauge field. 
We can choose $A^{t}_{\ell} = A_{3\ell} =0$ gauge for the non-zero ($\ell \neq 0$) modes, but we cannot choose this gauge for the zero modes of $A^{t}$. Usually the zero modes associated with $A^{t}$ (or $A_{3}$) can be studied by choosing an alternate gauge. Here, we use the first order formalism to treat the zero modes of the gauge field. In the first order formalism, the zero modes of $A_{i}$, $A_{t}$, $E_{3}$ and  $E_{i}$ are the degrees of freedom. We will call the zero modes as $A_{i}$, $A_{t}$, $E_{3}$ and  $E_{i}$ instead of introducing new symbols. Not all of these are dynamical degrees of freedom.   
After integrating out all the massive modes and after dimensional reduction, (\ref{Th1}) simplifies to
$$S \equiv  \int dt d^{2}x  \Bigg[{1\over 2g'_{1}} \mbox{tr} \left( E_{i}D_{t}A_{i} + A_{t}D_{i}E_{i}\right) + {1\over 4g'^{2}}\mbox{tr}\left(F_{ij}F^{ij}\right)  +  \lambda_{1}' \mbox{tr}\left([E_{i},E_{j}]^{2}\right)+ i\kappa' \mbox{tr}[E_{i},E_{j}]F^{ij} $$
 \begin{equation}
\label{Th2}
   + {1\over 2 {g'}_{3}^{2}}\int d^{2}xdt\mbox{tr}\(E_{3}^{2}\) +{1\over 2 \alpha^{2}}\int d^{2}xdt\mbox{tr}\left(\left(D_{i}E_{j}\right)^{2}\right)+ Q \int \mbox{tr} \(A \wedge F\)  + \mbox{irrelevant terms}\end{equation}
Note that the couplings get corrected after the massive modes are integrated out. Further, we can see that $E_{3}$ is not dynamical and the equation of motion for $E_{3}$ is $E_{3} = 0$! Hence, we can eliminate $E_{3}$ from the action. After eliminating $E_{3}$ we see that the action is same as the action for non-Abelian LCS theory. Note that this theory enjoys $z=2$ classical scaling symmetry when $x_{3}$ is compact, $$[t]=-2, [x_{i}] =-1, [E_{i}] =1, [A_{i}] =1, [A_{t}] =2.$$
Further, Galilean invariance is broken even when $r_{0} \rightarrow \infty$. This is due to the presence of dimension 6 operator. When $Q\neq 0$, it is not possible to scale $A$ to make $g_{1} =g_{2}$. However, this is possible when $Q=0$.
Note that $\alpha$ is a function of $m_{X}$, $L_{3}$, $\kappa$ and $\kappa_{6}$ {\ie} 
$$ {1\over \alpha^{2}} \sim \left({\kappa \over \kappa_{6} {\cal M}^{2}(m_{X}, L_{3})}\right) $$ 
where ${\CM}^{2}$ is a mass scale that appears in the action after integrating out the massive modes, 
which
is therefore
a function of $m_{X}$ and $L_{3}$.
\coo{Note that in \eqref{Th2}, we assumed that the coefficient of $\tr E_j^2$ is tuned to zero when
integrating out the massive modes.}

\section{The dependence of the gap on the CS level}
\label{sec:gap}

We have argued that the field theory dual of (\ref{Lifcut}) is a non-Abelian Lifshitz-Chern-Simons theory.
Our gravity description ceases to exist when the Chern-Simons level is turned off. 
Reference \cite{MikeNA} shows (perturbatively) that the weakly coupled theory 
without CS term flows to a free theory in the IR. Though this need not be true if we start the flow at strong coupling, this suggests that a classical supergravity description of the groundstate should not exist when $Q=0$. 

When $Q\neq0$, our gravity solution (terminated at $r_\star$) has a minimum value of the warp factor,
indicating that the mass gap has a non-trivial dependence on the Chern-Simons level. 
We will now show this more explicitly by computing masses of scalar glueballs. 
Note that parity $P$ is not a good quantum number as it is explicitly broken by the CS term. The fluctuations of dilaton, axion and $g_{33}$ mix as they are dual to gauge invariant operators with the same quantum numbers. The mixing between $\delta g^{\alpha}_{\alpha}$ and the dilaton 
(or the other two modes) is suppressed in the large $N$-limit (see \cite{Csaki}). These modes cannot mix with any other fluctuation as other modes have different quantum numbers ($J$ and $C$). 

We can compute the $0^{+}$ glueball spectrum by solving for the linearized fluctuations of the dilaton, axion and $g_{33}$ subject to regular boundary conditions at $r=r_{\star}$ and the UV normalizability condition. At first glance, this might seem unreasonable as the metric is geodesically incomplete if $r$ is restricted to lie within $r_{\star}$. However, the geodesics carrying zero momentum along $x_3$ do not penetrate the region past $r_{\star}$. {\footnote{ This is clear from the computation in \cite{Copsey}.}}  
The solution we find in the section \ref{sec:rep2} will reveal that the calculation
of this section is a good approximation in the regime $r_\star^{-1} \ll L_3^{-1}$.

The frequency of the $2+1$ dimensional theory is obtained by scaling the frequency of $3+1$ dimensional theory by $L_{3}$.{\footnote{Recall that we had scaled by a factor of $L_{3}$ in the dual field theory to absorb an overall factor of $L_{3}$ after dimensional reduction.}} We will 
only consider the modes with zero spatial momentum, since we are only interested in the masses of the glueballs. Note that the metric has no explicit depence on $x_{3}$ and only derivatives of the axion field can appear in the equations of motion of the scalar field (dilaton, axion and $g_{33}$) fluctuations. Hence, the equations of motion for these fluctuations cannot have explicit dependence on $x_{3}$. This implies that the fluctuations with zero momentum along $x_{3}$ get decoupled from the non-zero modes.{\footnote{The equation is in the variable separable form and hence the modes with non-zero momentum along $x_{3}$ can be separated from the zero mode.}}  Let us now choose the following ansatz for the fluctuations of dilaton, axion and $g_{33}$
\be \delta \Phi = \varphi(r) e^{i \omega t/L_{3}  },
~~ \delta \chi = \chi_{1}(r)e^{i \omega t/L_{3}  }, 
~~\delta g_{33} = {h_{33}(r)\over r^{2}} e^{i \omega t/L_{3}  } ~.\ee
Here $\omega$ is the frequency in the $2+1$ dimensional theory. 
We define 
\be \label{eq:gammadef}\Gamma = {Qe^{2\Phi_{0}}\over \sqrt{2}L_{3}}. \ee
The contribution of the factors in the metric to the mass gap can only be functions of $\Gamma$ and $r_{\star}$. This will allow us to study contribution of the axion to the dependence of the 
mass gap by just studying the dependence of the mass gap on $Q$.  
The equations  of motion for the linearized fluctuations ($\varphi$, $\chi_{1}$, and $h_{33}$) are
\be e^{2 \Phi_{0}} Q \chi_{1} = \omega h_{33} \ee
\be e^{- 2\Phi_{0}}  r \omega^{2} \(- 8 r_{\star}^{2} h_{33} + r^{2}\(r^{2} - r_{\star}^{2}\) \Gamma^{2} \varphi\) + {4Q^{2} \over \Gamma^{2}} r^{2} (3\varphi_{1}' - r \varphi'') = 0 \ee
\be  e^{- 2\Phi_{0}}  r^{3} \(r_{\star}^{2} - r^{2}\) \omega^{2} h_{33}(r) + 4 {r_{\star}^{2} Q^{2} \over \Gamma^{4}} \(2 r \alpha^{2} \varphi(r) - 3 h_{33}'(r) + r h_{33}''(r)\) = 0~.\ee
Note that the first equation was used to eliminate $\chi_{1}$ from the other two equations of motion. The masses of the glueballs are eigenvalues of the above equations subject to regularity condition at $r=r_{\star}$. Further, the glueballs correspond to the normalizable modes of these fluctuations and hence the modes must satisfy normalizability condition. We can see that for fixed $\Gamma$ and $r_{\star}$,  the mass gap will have non-trivial dependence on $Q$.{\footnote{Fixing $r_{\star}$ is analogous to specifying the confinement scale in 3+1 dimensional YM theory. This scale is generated from a dimensionless coupling by dimensional transmutation.}}

\begin{figure}[h]
   \centering
   {\includegraphics[scale=0.75]{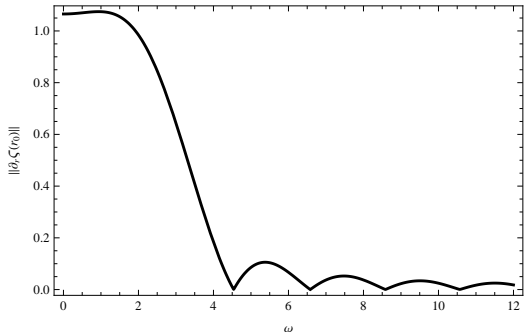} 
   }
   \caption{\label{fig:one}A graphical approach to find the glueball masses. We have chosen $r_{\star} =1$ and $\Gamma=10$.}
\end{figure}

\begin{figure}[h]
   \centering
   {\includegraphics[scale=0.75]{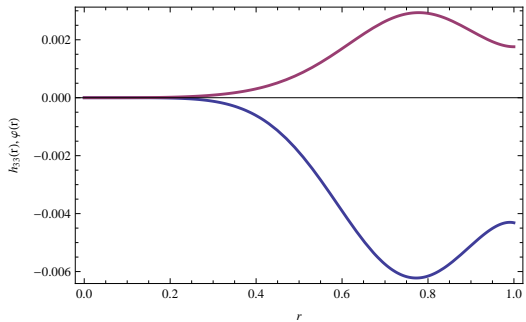} 
   }
 \caption{
   \label{fig:two}
  Plot of the normalizable mode of $g^{33}, \chi$ for the lowest eigenvalue. Note that the fluctuations of the axion is proportional to $h_{33}$ and hence it is not plotted here.}
 \end{figure}

The above equations can be solved numerically by shooting. Here, we integrate from the boundary to the infrared by specifying normalizable boundary conditions for the fluctuations and using the shooting method to satisfy the regularity condition: $||\zeta'(\rs)||=0$ where $ \zeta = [h_{33}, \varphi, \chi_{1}]
$. In order to specify the boundary condition we assume a power series expansion around $r=0$ for the fluctuations and determine the coefficients (up to four terms) for which the modes are normalizable and the equations of motion is satisfied approximately near the boundary. We see that, $\chi_{1}$ must fall of as $r^{6}$ near the boundary for all modes to be normalizable. With these boundary conditions, we integrate the system of equations numerically and determine the values of $\omega$ for which the regularity condition is satisfied. Figure \ref{fig:one} shows a plot of $||\zeta'(\rs)||$ as a function of ${\Omega}  = \omega e^{-\Phi_{0}}/Q$ for $\Gamma = 10$ and $r_{\star} = 1$. The points at which the graph touches the $\omega-$axis are points at which the regularity condition is satisfied. We can see from the figure that for $\omega \approx e^{\Phi_{0}} Q  \{4.5,~ 6.5,~ 8.5 ,~ 10.5 ,..\}$ we get normalizable solutions that satisfy regularity boundary condition. These are the values of the glueball masses measured in units where $r_{\star} = 1$. Figure \ref{fig:two} shows the radial profile of the solution corresponding to the lowest eigenvalue ($\Omega \approx 4.5$).  

We emphasize that the identification of parameters 
between bulk and boundary
described above is subject to renormalization.
Further, the overall normalization of the couplings is difficult to obtain without 
further microscopic information.
The dependence of the mass gap in the gauge theory found above
is obtained by fixing $\Gamma$ and $r_{\star}$. 
The logic is that $\Gamma$ determines the coefficient $\kappa$ 
(as explained in section \ref{sec:qft}), while $\rs$ is analogous
to $\Lambda_{QCD}$ in QCD, \ie~a scale which determines the gauge coupling by dimensional
transmutation.  The CS coefficient then maps directly to the axion slope.

We close this section with some comments.
\begin{enumerate}

\item Our results suggest that 
turning on a Chern-Simons term
in Lifshitz gauge theory changes the sign of the beta functions
computed in \cite{MikeNA}, and leads to a gapped state.
This is a counterintuitive claim\footnote
{We thank Mike Mulligan for emphasizing this point to us.}.
Adding a CS term to 
an ordinary ($z=1$) gauge theory in 2+1 dimensions, Abelian or non-Abelian,
weakens the long-range gauge dynamics. 
This is simplest to see in the (gaussian) Abelian
Maxwell-Chern-Simons theory 
(with noncompact gauge group)
where the CS term produces a mass for the gauge boson 
\cite{Deser:1982vy}.
In the non-Abelian theories or in the compact U$(1)$ gauge theory, the story is more subtle,
but the conclusion is the same
\cite{Affleck:1989qf, Rey:1989jg, Grignani:1996ft}.
A recent paper which relies crucially on this effect is \cite{FCI}.

Recent studies of Abelian Lifshitz-Chern-Simons \cite{Mike}
make it clear that intuitions from $z=1$ gauge theories do not always apply
to Lifshitz gauge theories.  
It appears that, the long-range gauge dynamics 
in the model
\eqref{Th2}
is a complex interplay between the parameters that we call as $g_1'$, $\lambda'$ and the 
Chern-Simons level. 
The $\lambda'$ term is irrelevant in the $z=1$ case. 
Another crucial difference is that a $z=1$ theory contains a $E_i^2$ term while it is absent in LCS theory.     
The perturbative dynamics of this model should be analyzed.

\item It would be interesting to calculate 
on both sides of the duality proposed in this paper
observables which are sensitive to the 
Chern-Simons level $Q$.
An interesting class of examples is given by Wilson-'t Hooft loops.
In perturbation theory around the gaussian model, the CS coupling has an immediate
effect on Wilson loops via its influence on the gluon propagator.
At strong coupling, the effect of the axion profile is more subtle.
In the bulk, the area of a fundamental string 
worldsheet ending on the quark trajectory computes the 
expectation value of a Wilson loop \cite{Maldacena:1998im, Rey:1998ik}.
But a fundamental string does not couple to the axion profile,
and its action only sees the axion slope through the (weak) metric 
dependence on $Q$.
In contrast, a D-string does couple to the axion, via the worldvolume Chern-Simons term
\be S_{\text{D-string}} ~\contains \int_{\text{D-string}} \chi F ~,\ee
where $F$ is the worldvolume gauge field on the D-string.
A D-string configuration with $p$ units of worldvolume flux 
carries F-string charge $p$, and therefore computes
a mixed Wilson-'t Hooft loop describing
the holonomy for a $(p,1)$ dyon.
This apparent tension is another counterintuitive 
manifestation of the effects of the CS level in the Lifshitz gauge theory.

\item We should comment on what happens to our theory when the hierarchy in Fig.~\ref{fig:scheme}
is re-ordered. 
If the radius $L_3$ is smaller than the deformation scale $\Gamma$, 
our argumentation in section \ref{sec:qft} breaks down,
because it relied on supersymmetry to make the identifications of the deformations.  
The gravity solution remains regular, however.
Given the definition \eqref{eq:gammadef}
of $\Gamma$,
we note that $\Gamma \gg L_3^{-1}$ (at weak string coupling) requires $Q\gg1$.

If the radius $L_3$ is taken larger than the IR scale $r_\star$,
then the model describes a 3+1 dimensional field theory 
with explicitly broken translation invariance (by the axion profile);
the fact that the $x_3$ circle becomes timelike for $r > r_\star$ 
now represents a more serious problem, and the reader is referred to \S\ref{sec:repair}.

\item Since it does not rely on the structure of the $S^5$, 
the null deformations of $AdS_5 \times S^5$ described above
have a generalization to many known $AdS$ vacua of supergravity.

\item 
From the gravity solution, we see that 
the IR scale $r_\star$ and the KK scale $L_3$ 
may be made arbitrarily different while maintaining control over the solution.
Why does this solution allow for such a parametric
separation?
A simple QFT answer to this question would be progress toward
a solution of confinement, but we offer the following observations.
The reduced symmetry of the problem --
$P$ and $T$ violation as well as Lorentz violation -- 
allows for new ingredients, which are apparently helpful for this purpose.
On the one hand, the $P$- and $T$-violating axion gradient along the circle
provides an energetic incentive for the radius of the circle
not to shrink.
On the other hand, the Lorentz-violating couplings 
of the gauge fields are dual to the exotic boundary conditions on the vielbein;
these boundary conditions play an important role in determining the IR scale.

\item Recall that $\tr\(E_i^2\)$ is a relevant operator in LCS theory. 
At least in the Abelian case, when the LCS theory is deformed by this operator (with a positive coefficient), it flows to a theory with $z=1$ scaling symmetry.  
When the coupling to this relevant operator is negative, rotational symmetry is spontaneously broken  \cite{Mike}. Can such a nematic phase be seen in the gravity dual\footnote
{We thank Shamit Kachru for asking this question.}? 
At present, we do not have a concrete answe, but we 
give some preliminary ideas for understanding the relevant deformation in the gravity dual. 
We proceed by noting that the action of the transformation $t \rightarrow t + s x_3$ on DLCQ of \eqref{Th1} describes a deformation of \eqref{Th1} by $\tr\(E_i^2\)$ term (with a coefficient proportional to $s$). The theory obtained by compactifying $x_3$ (after the transformation) is no longer invariant under $z=2$ scaling symmetry. The theory obtained by reducing along $x_3$ also contains the terms $\tr\(E_i^4\)$, $\tr\(E_i^2\)^2$ and other terms that are irrelevant with respect to both $z=1$ and $z=2$ scaling. On the gravity side, the transformation $t \rightarrow t + s x_3$ generates a new solution with the metric given by 
$$ ds^2_{new} = L^{2}\({2 dx_{3} dt + d\vec x^{2} + dr^{2} \over r^{2}} +\({s\over r^2} + {Q^2 e^{2 \Phi_0} \over 4 L_3^2} - {r^2 \over r_0^4}\) dx_{3}^{2} \) $$
A nematic phase would be encouraged when the coefficient of $\tr\(E_i^2\)$ is negative. 
This happens when $s<0$ in which case $x_3$ is a compact time-like direction. 
It seems that this particular holographic realization of Lifshitz Chern-Simons theory 
does not admit a description of the nematic phase within the gravity regime.
\end{enumerate}

\section{Resolution of the ``hidden singularity''}
\label{sec:repair}

  If the geometry (\ref{Lifcut}) is extended past $r = r_\star$, the $x_3$ circle 
becomes timelike \cite{Myers, Copsey}.
The presence of CTCs in the bulk need not be related to violation of unitarity in the UV description of the dual field theory.{\footnote{In some examples of rotating black holes in 2+1 D\cite{Chronology}, the presence of CTC region was attributed to the violation of unitarity bound in the dual field theory. In these examples, either the null-energy condition is violated or the asymptotic geometry contains a region with CTCs.}} 
Note that the asymptotic geometry of (\ref{Lifcut}) is free of CTCs and the matter supporting this metric does not violate null energy condition (even in the region with CTCs). Hence, the CTC region is not linked to violation of unitarity in the UV description. 
Rather, the existence of the CTC region is an indication of IR instability in the state of the dual field theory.
The local mass${}^2$ of KK modes become tachyonic: $m^2(r)_{p^3} = g^{33} p_3^2 < 0 $.  Further,
wound strings may become tachyonic when the radius is of order of the string length $\sqrt{\alpha'}$;
their condensation would excise the CTC region \cite{AdamsMaloney, Costa:2005ej}.

We will now present two different resolutions of this singularity and discuss their implications for the IR behavior of the dual field theory.

\subsection{Running Dilaton}
\label{sec:rep1}

In this section we show that it is possible to resolve the singularity by imposing non-trivial boundary conditions on the dilaton. As discussed earlier, such non-trivial boundary conditions on the dilaton corresponds to deformation of the dual field theory by a dangerous irrelevant operator (see appendix A). We will see that for one 
sign for the dangerous irrelevant coupling, the IR singularity is resolved.  
Note that this is the groundstate of a {\it different} theory from that determined by the asymptotics
of \eqref{Lifcut} -- the dilaton boundary conditions indicate a perturbation
of the dual QFT by a (dangerously-irrelevant) operator.

The solution with a running dilaton is given by,
\be ds^{2} = L^{2}\({2H_{d}(r)  dx_{3} dt + d\vec x^{2}  \over r^{2}}+ {dr^{2}\over r^{2}H_{d}(r)^{2}}+f_{d}(r) dx_{3}^{2}  + ds^{2}_{\Omega^{5}}\), ~H_{d}(r) = \sqrt{1-{r^{4}\over r_{1}^{4}}} ~~~\label{Lifdil} \ee
$$F_{5} =2L^{4}(1+ \star)\Omega_{5} , ~~ C_{0} = {\Qa x_{3} \over L_{3}}, ~~ \Phi = \Phi_{0} + \log \(H_{d}(r)\), ~~ f_{d}(r) =  {H_{d}(r)\over r^{2}}{\Qa^{2} e^{2\Phi_{0}}  \over 4L_{3}^{2} } {\cal F}\(\sin^{-1}\({r^{2}\over r_{1}^{2}}\) \)  $$       
where,
$$ \CF(x) =  r_{1}^{2} x + r_{1}^{2} \int_{0}^{x} \xi \tan \xi d\xi - {r_{1}^{4}\over r_{0}^{4}} \log\(\cos^{2} x\) $$
The integral in ${\CF}(x)$ can be evaluated in terms of Polylogarithms. It can be checked that this solution approaches (\ref{Lifcut}) when $r_{1} \rightarrow \infty$ and has the same asymptotic behavior as (\ref{Lifcut}),
except for the dilaton profile. 
In this solution, the geometry ends at $r=r_{1}$ and $f_{d}(r)$ remains positive throughout the geometry. 

There is a curvature singularity at $r=r_{1}$ which can be resolved by uplifting the solution to 11D supergravity.{The details of the uplifting procedure is discussed in appendix B} In the following, we will show that this geometry is geodesically complete. It is sufficient to focus on the case of null geodesics. The null geodesic equation is given by
$ \dot{r}^{2} + V_{eff}(r) = 0$ where 
 $$V_{eff}(r) =r^{2} H_{d}^{2}\( -E^{2} {\Qa^{2} e^{2\Phi_{0}}  \over 4L_{3}^{2} } {\cal F}\(\sin^{-1}\({r^{2}\over r_{1}^{2}}\) \){ r^{4}\over H_{d}(r)} - r^{2} {2 p_{3}E\over H_{d}(r)}   +r^{2}\({p_{1}^{2} + p_{2}^{2}}\) \)$$
In the above expression $E$, $p_{3}$ and $p_{i}$ are the conserved quantities associated with $\del_{t}$, $\del_{3}$ and $\del_{i}$. Note that $V_{eff}(r_{1})$ is zero and hence the maximum possible value of $r$ for radially ingoing geodesics is $r_{1}$. Hence, no geodesic can penetrate into the region past $r_{1}$. Hence, the 11D solution obtained by uplifting the solution in (\ref{Lifdil}) to 11D supergravity is regular and provides a resolution of the ``hidden singularity''. 
There is also another solution with running dilaton obtained reversing the sign of the dilaton gradient near the boundary {\ie}, $\Phi = \Phi_{0} - \log H_{d}(r)$. In this case, $g_{33}$ vanishes before $H_{d}(r)$ vanishes and the geometry is geodesically incomplete (or contains regions with CTCs). Hence, we did not present this solution here. 

The dual field theory interpretation of the resolution discussed in this subsection is the following. Turning on a dangerous irrelevant deformation causes the gauge coupling to run. The gauge coupling can become strong or weak in the IR depending on the sign of the dangerous irrelevant deformation (dilaton gradient at the boundary). When the coupling becomes weak in the IR, the dynamics is controlled by the CS term which leads to an IR instability. When the gauge coupling becomes strong in the IR, the theory confines and overrides the effect of the CS term. The confinement scale is not related to the axion flux or Chern level in this solution. In the next section we will provide an alternate resolution of the singularity. 

\subsection{Excision of the CTC region}
\label{sec:rep2}

In this subsection, we provide an alternate resolution of the singularity which is similar to the enhan{\c{c}}on mechanism \cite{Clifford}. 
Unlike the enhan{\c{c}}on, there is no enhancement of gauge symmetry. Closed time-like curves can be prevented by placing localized sources at $r=r_{\star}$. This will preserve the asymptotic form of the metric but modifies the region beyond $r_{\star}$. We will show that the domain wall is described by smeared D3 branes located at $r=r_{\star}$. In a supersymmetric theory, the location of the brane specfies the 
vacuum expectation value (VEV) for some scalar field in the dual gauge theory. The moduli space of the theory describes all possible locations of the branes. In the system described here, 
the location of the branes is uniquely specified by the asymptotic boundary condition. 
This is natural in a theory with broken supersymmetry -- the moduli space is lifted,
leaving a unique groundstate.

Unlike in the previous subsection, the solutions described here are dual to states of 
the same QFT as \eqref{Lifcut}, and hence represent a possible endpoint of
the localized instabilities associated with the CTC region.

Before we describe this resolution, let us remind ourselves about the gravity dual of a spherically symmetric shell of D3 branes (smeared over $S^5$) that describes a special point in the Coulomb branch of $\CN =4$ SYM theory. 
The relevant solution of IIB supergravity is
\begin{eqnarray} 
ds^2 = {L^2\over r^2} \(\eta_{\mu\nu} dx^{\mu}dx^{\nu} + dr^2 + r^2 ds^2_{\Omega_5} \), 
~~F_5 = 2L^4 (1+\star)\Omega, &
 & \text{for}~ r<r_{\star}\cr\cr
 ds^2 = {L^2\over r_{\star}^2} \(\eta_{\mu\nu} dx^{\mu}dx^{\nu}\) + {L^2 r_\star^2 \over r^4}\(dr^2 + r^2 ds^2_{\Omega_5} \), ~~~~~ F_5 =0,&
  & \text{for}~ r>r_{\star}
\label{eq:shell}
 \end{eqnarray}
 where $\Omega$ is the volume form on the 5-sphere.
The interior geometry ($r > r_{\star}$) is just flat space{\footnote {The solution looks more familiar in the coordinate system where boundary is at infinity ($r=L^{2} \rho^{-1}$, $r_{\star}=L^{2} \rho_{\star}^{-1}$). In this coordinate system, the solution is given by
$$ ds^2 = {\rho^2\over L^2} \(\eta_{\mu\nu} dx^{\mu}dx^{\nu} \)+ {L^2 \over \rho^2} \(d\rho^2 + \rho^2 ds^2_{\Omega_5} \), \quad \text{for}~ \rho>\rho_\star$$
$$ ds^2 = {\rho_\star^2\over L^2} \eta_{\mu\nu} dx^{\mu}dx^{\nu} + {L^2 \over \rho_\star^2}\(d\rho^2 + \rho^2 ds^2_{\Omega_5} \), \quad \text{for}~ \rho<\rho_\star$$
In this coordinate system the boundary is at $\rho = \infty$.}} and the exterior geometry ($r<r_{\star}$) is $AdS_5 \times S^5$. The D3 branes are localized around $r=r_{\star}$ and acts as the source for Israel stress tensor ($S_{\mu\nu}$). Note that the metric is continuous at $r=r_\star$ and we will now show that it also satisfies the Israel junction condition. The junction stress tensor is given by
\be K^+_{\mu\nu} + K^-_{\mu \nu} - G^{jun}_{\mu \nu}\(K^+ + K^-\) = {-{2\over L}G^{jun}_{\mu \nu}} = S_{\mu\nu} , \quad \text{for}~\mu,\nu \in 0,1,2,3 \label{isr1}\ee
\be K^+_{ij} + K^-_{ij} - G^{jun}_{ij}\(K^+ + K^-\) = {-{2\over L}G^{jun}_{ij}} = S_{ij} , \quad \text{for}~i,j \in S^{5} \label{isr2}\ee
where $K^{\pm}_{AB} = \mp {1\over 2} n^{r}\del_{r} G^{jun}_{AB}$ and $G^{jun}_{AB}$ is the induced metric at the junction. The integrated Einstein equation tells us that the the Israel stress tensor is sourced by the D3 branes {\ie}
\be S_{AB} = {N\over \text{Vol}(S^{5})} {\delta S_{D3} \over \delta G^{jun}_{AB}} \label{junccond}\ee
The right hand side of the above equation is the stress tensor of $N$ D3 branes smeared over $S^{5}$ located at $r_{\star}$ and $S_{D3}$ is the world volume action of $D3$ branes.
 The worldvolume action for D$p$ branes (with world volume gauge fields set to zero) is
$$S_{Dp} = -T_{Dp}  \int d^{p+1}\xi e^{{p-3\over 4}\Phi}\sqrt{G_{Einstein}^{\text{PB}}}  + S_{WZW}$$
where $G^{PB}_{Einstein}$ is the D$p$ brane metric in Einstein frame. Note that when $p=3$ there is no source term for the dilaton and the dilaton remains constant. After taking the derivative of the worldvolume action with respect to the metric, we can see that the Israel junction conditions in (\ref{junccond}) are satisfied.  

When one of the space directions of $AdS$ 
is compactified 
with APBC on the fermions around this compact direction
Horowitz and Silverstein \cite{Horowitz:2006mr} argued that the above solution is unstable and decays to an
AdS soliton.  Since the interior geometry is flat, perturbative techniques can be employed to show the existence of 
closed string tachyons from strings winding around the compact direction. 
Tachyon condensation excises the IR region leaving behind a cigar shaped geometry reflecting the confining nature of 3D Yang-Mills theory. 

We may also expect such tachyons to develop in the region surrounding a region of CTCs.
Let us make use of the intuition we get from Maxwell-Chern-Simons theory to guess the correct IR behavior of the solution in (\ref{Lifcut}). Turning on a CS interaction weakens the gauge dynamics in the IR and can prevent confinement. Confinement in the dual gauge theory is prevented if the D3 brane shell system is stable in the presence of a linear axion profile. In the following, we will show that region with CTCs in (\ref{Lifcut}) can be removed by placing a shell of D3 branes at $r=r_{\star}$.

The solution describing the shell of D3 branes in the presence of axion flux is
$$ ds^{2}_{\text{ext}} = L^{2}\({2 dx_{3} dt + d\vec x^{2} + dr^{2} \over r^{2}} +f_{ext}(r) dx_{3}^{2}  + ds^{2}_{\Omega^{5}}\), \quad f_{ext}(r) =  {\Qa^{2} e^{2\Phi_{0}}  \over 4L_{3}^{2} }\(1 - \left({r^{2} \over r_\star^{2}}\right)\right)$$ 
\be F_{5} =2L^{4}(1+ \star)\Omega_{5} , \quad C_{0} = {\Qa x_{3} \over L_{3}}, \quad \Phi = \Phi_{0}, \quad \text{for}~r<r_{\star} \label{Lifenhan} \ee
$$ ds^{2}_{\text{int}} = {L^2\over r_{\star}^2} \(2 dx_{3} dt + d\vec x^{2}+f_{int}(r)dx_{3}^{2} \) + {L^2 r_\star^2 \over r^4}\(dr^2 + r^2 ds^2_{\Omega_5} \)  \quad f_{int}(r) =  {\Qa^{2} e^{2\Phi_{0}}   r_{\star}^{2}\over 12 L_{3}^{2} }\( {r^{4} \over r_\star^{4}}- {r_{\star}^{2} \over r^{2}}\right)$$
$$F_{5} =0 , \quad C_{0} = {\Qa x_{3} \over L_{3}}, \quad \Phi = \Phi_{0}, \quad \text{for}~r<r_{\star}  $$
There is no jump in axion flux and hence D7 brane sources are absent in this solution. The jump in 5-form flux is sourced by the shell of D3 branes. Note that the metric is continuous at $r=r_{\star}$.  We will now show that this solution also satisfies Israel jump conditions, if $r_\star$ has the same relation to the 
UV variables as previously \eqref{eq:rstar}.
The junction stress tensor is
$$ K^+_{\mu\nu} + K^-_{\mu \nu} - G^{jun}_{\mu \nu}\(K^+ + K^-\) = {-{2\over L}G^{jun}_{\mu \nu}} = S_{\mu\nu} , \quad \text{for}~\mu,\nu \in 1,2$$
$$ K^+_{ij} + K^-_{ij} - G^{jun}_{ij}\(K^+ + K^-\) = {-{2\over L}G^{jun}_{ij}} = S_{ij} , \quad \text{for}~i,j \in S^{5}$$
$$K^+_{33} + K^-_{33} - G^{jun}_{33}\(K^+ + K^-\) = 0 = S_{33}$$ 
$$ K^+_{t3} + K^-_{t3} - G^{jun}_{t3}\(K^+ + K^-\) = {-{2\over L}G^{jun}_{t3}} = S_{t3} $$
Since $G^{jun}_{33} = 0$ at $r= r_{\star}$, we can write $S_{33}$ as $-(2/L) G_{33}^{jun}$. Hence, the form of the Israel stress tensor is same as (\ref{isr1}) and (\ref{isr2}). We already saw that a shell of D3 branes can provide this stress tensor. Hence, this solution provides a consistent way of removing the region with closed time like curves. 

In the solution \eqref{Lifenhan}, 
the IR geometry is a planewave.
Tidal forces become large as $r\to\infty$.
This sort of mild singularity is familiar from the Lifshitz solution
and we regard it as physically acceptable.

Evaluating the regulated on-shell action of the solution
\eqref{Lifenhan}, we find that it compares favorably to that of 
\eqref{Lifcut}.  We conclude that the solution \eqref{Lifenhan} is a truer groundstate.
We cannot exclude the possibility of more favorable solution, 
such as a smooth solution which terminates at a finite value of $r$.

Just as the D3-brane shell solution \eqref{eq:shell} exhibits a mass gap
in the spectrum of single-trace operators, 
so will \eqref{Lifenhan}.  The reason in both cases is that the 
5-sphere shrinks at $r=\infty$ -- the geometry for $r > r_\star$ 
is roughly a (compact) ball.
The analysis of \S\ref{sec:gap} becomes a good approximation
when this ball is small: $r_\star \gg  L_3$.
We must leave an analysis of the spectrum in the general case 
for the future.


\vskip.2in
{\bf Acknowledgements}

We thank 
Ben Freivogel, Sean Hartnoll, Shamit Kachru, Rob Myers, Eva Silverstein, Brian Swingle, Allan Adams, Tom Faulkner, Nabil Iqbal, Hong Liu, Vijay Kumar, Daniel Park, Krishnan Narayan, Sho Yaida
and especially Mike Mulligan
for discussions, comments and encouragement.  This work was supported in part by
funds provided by the U.S. Department of Energy
(D.O.E.) under cooperative research agreement DE-FG0205ER41360, 
in part by the Alfred P. Sloan Foundation,
and in part by the National Science Foundation under Grant No. NSF PHY05-51164.

\appendix
\section{Dilaton-driven confinement}
\label{sec:dilatondriven}

Here we review a holographic model for confinement studied in \cite{Gubser}.\footnote
{This section contains some new results, some remarks benefitting from a decade of hindsight, and some minor differences in the style of presentation.} 
As in \cite{Witten}, there is a circle with APBCs.  However, the confinement scale is not determined by the UV radius of the shrinking circle but is determined from the boundary conditions on the supergravity fields. The contents of this section is somewhat disconnected from the rest of the paper. The results will be helpful in section 4. 

Gubser \cite{Gubser} found (numerically){\footnote{The analytical solution has appeared previously in \cite{Kehagias}. 
}}
an asymptotically $AdS_5 \times S^5$ solution solution of type IIB supergravity
with unusual boundary conditions for the dilaton field.
 The resulting non-trivial profile for the dilaton leads to confinement. The following is the solution that was studied in \cite{Gubser}
$$ ds^{2} = L^{2}\(1- {r^{8} \over r^{8}_{0}}\)^{1/2} \( {-d\tau^{2} + dy^{2} + d\vec{x}^{2} + dr^{2} \over r^{2}}\) +L^{2} ds^{2}_{S^{5}} $$
\be F_{5} = L^{4}\(1 + \star \Omega_{5}\), \Phi = {\sqrt{6} \over 2} \log\({r_{0}^{4} - r^{4}\over r^{4}+  r^{4}_{0}}\) \label{gubsol}\ee
where $F_{5}$ is the RR five-form flux of Type IIB supergravity and $\Phi$ is the dilaton. Here, we will assume that $y$ is a compact direction with period $L_{y}$. The equations of motion and other details about the solution can be found in appendix~\ref{app:solutions}. 
This appendix also contains a family of solutions of which the above solution 
is a special member
distinguished by the fact that it preserves Lorentz invariance in the UV. 
The dilaton becomes singular at $r=r_{0}$. There is also a curvature singularity. However, the metric is conformal to a regular metric. In fact, it is possible to resolve the singularity by ``uplifting'' the solution to a regular solution of 11D supergravity (see appendix~\ref{app:solutions}). 

The fact that the geometry ends smoothly in the IR signals a mass gap in the dual field theory\footnote{This solution is only relevant if fermions satisfy anti-periodic boundary conditions around $y$.}.
In particular, this indicates that the matter fields have been made massive. Further, the presence of $S^{5}$ factor in the solution implies that this mass is $SO(6)$ invariant. In the $AdS$ soliton case masses for all the fields are generated by the boundary conditions on fermions. In the present case, there must be two different mass scales. 

Note that we can give the scalars an $SO(6)$-invariant mass by adding 
\be\label{eq:massterm} L_m = m_X^2 \tr \sum_{I=1}^6 X_I^2\ee 
to the $\CN = 4$ SYM Lagrangian. The fermions get mass through anti-periodic boundary condition. It was suggested that the mass term of the scalar ($m_{X}$) is responsible for the non-trivial behavior of the dilaton in the bulk.  In particular, it was argued that the non-trivial boundary condition on the dilaton is induced by a ``string field'' that is dual to ${\cal O}_{K} = m^{2}_{X} \mbox{tr}\left(X^{I}X^{I} \right)$. Although ${\cal O}_{K}$ is a relevant operator at weak coupling, 
it acquires a large anomalous dimension at strong 't Hooft coupling and hence it is not 
visible in supergravity. 
Rather, it is dual to an excited mode of the IIB string in $AdS_5 \times S^5$ of mass of order $\sqrt{\lambda}$.
Because of its large mass, this ``string field'' has a profile that decays extremely rapidly 
near the UV boundary of $AdS$.
Hence, the effect of the ``string field'' 
on supergravity fields is felt just near the boundary. 
This effect appears as a non-trivial boundary condition on the dilaton. 
Such a boundary condition on the dilaton 
can in turn be described as a large-dimension multi-trace deformation of the QFT action\cite{Witten:2001ua, Berkooz:2002ug}.
From the fact that such an irrelevant operator has an important effect on the IR physics, 
we are forced to call it `dangerously irrelevant'.

It is difficult to justify the previous statements rigorously. In \cite{Gubser}, the following heuristic calculation was presented to justify this picture and to estimate the mass gap in terms of $m_{X}$. 
Our purposes in redoing this calculation here are twofold:
\begin{enumerate}
\item to make explicit the dependence of the IR scale $r_0^{-1}$ 
on non-normalizable deformations near the UV boundary,
\item to interpret the holographic renormalization for the dilaton field $\Phi$ 
in this context in terms of a boundary potential.
\end{enumerate}

Assume that adding the mass term 
\eqref{eq:massterm}
for scalars to the ${\CN} =4$ Lagrangian (with UV cut-off  ${\CM}_{1} > m_{X}$)
corresponds to turning on a source for an excited string field 
 $\phi_{K}$ in the bulk (with cut-off $r=\varepsilon_{1}$). 
In this calculation $\phi_{K}$ is treated as a linear perturbation in the bulk with mass $m_{K} $. Let us assume that $\phi_{K}$ interacts with the dilaton through an interaction term (in the bulk Lagrangian) of the form $W(\Phi)\phi_{K}^{2}$.
(The choice of this coupling is made for convenience and should not be taken too literally.)
The equations of motion can be written as 
\be \nabla^{2} \phi_{k} + m^{2}_{K} \phi_{K} = 0 \implies \phi_K 
\buildrel{r \to 0} \over {\simeq } \mu^2 \left( {r \over \varepsilon_{1}} \right)^{4-\Delta_K}\ee
\be \nabla^{2} \Phi = W'(\Phi) \phi_K^2~.\ee
Note that $W(\Phi)$ does not affect the $\phi_{K}$ equation of motion since $m_{K}$ is very large. 
We must further specify boundary conditions
for the dilaton at the UV cutoff $\epsilon_1$:  $\Phi(\varepsilon_{1})= \Phi_{0}$ and $\del_{r} \Phi |_{r=\varepsilon_{1}} = 0$ and $\phi_K(\varepsilon_{1}) = \mu^2$. 

 Now, let us integrate out all modes with mass greater than $\CM_{2}$ (with ${\CM}_{2} < m_{X}, L_{y}^{-1}$) in the dual field theory. This corresponds to integrating along the radial direction from $\varepsilon_{1}$ to $\varepsilon_{2} (> \varepsilon_{1})$ in the bulk. Integrating the dilaton equation of motion once we get
\be {1 \over \varepsilon_{2}^3} \partial_r \Phi |_{r=\varepsilon_{2}} = \int_{\varepsilon_{1}}^{\varepsilon_{2}} 
    {d{r} \over {r}^5} \lambda W'(\Phi)\phi_K^2  
    \approx W'(\Phi_{0}) \mu^4 \int_{\varepsilon_{1}}^\infty {dr \over r^5}
     \left( {r \over \varepsilon_{1}} \right)^{8-2\Delta_K} 
     \ee
     \be
   \implies {1 \over \varepsilon_{2}^3} \partial_r \Phi |_{r=\varepsilon_{2}} = {W'(\Phi_{0})  \over 2\Delta_K - 4} {\mu^4 \over \varepsilon_{1}^4} \approx {W'(\Phi(\varepsilon_{2}))  \over 2\Delta_K - 4} {\mu^4 \over \varepsilon_{1}^4}
   \ee
This implies that the boundary condition on $\Phi$ at $r=\varepsilon_{2}$ is determined by $\mu/\varepsilon_{1}$ (assuming $W$ and $\Delta$ are known). Note that $m_{X} \sim \mu/\varepsilon_{1}$. We have to choose $\mu < 1$ for the dual field theory to make sense. Further since $\Delta_{K} \gg 1$, we can see that the confinement scale
$m_{\text{confine}} \sim 1/r_{0} < m_{X}$. 
The separation between $m_{\text{confine}}$ and $m_X$ 
depends on the precise form of $W(\Phi)$. When $W'$ is non-zero the boundary condition on the dilaton differs from that in the $\CN=4$ theory. 
As we describe in appendix~\ref{app:counterterms}, a well-defined variational principle
requires a boundary potential for the dilaton (this serves as the counterterm). 

We emphasize that pure $AdS$ does not satisfy this non-trivial boundary condition. 
The boundary potential for the dilaton 
represents a deformation of the $\CN=4$ theory, and 
cannot be interpreted as a parameter specifying
a state of the $\CN=4$ theory.
A more modern perspective 
on the effects of such a boundary potential for the dilaton 
was given in
\cite{Witten:2001ua, Berkooz:2002ug}:
$W(\Phi)$ encodes a deformation of the dual field theory action
by a combination of multitrace operators.
From this point of view, the calculation \cite{Gubser} 
that we have just done is a nice example 
of holographic Wilsonian RG described in \cite{Heemskerk:2010hk, Faulkner:2010jy}.
The precise relationship between the SO$(6)$-invariant
scalar mass and the multitrace operator encoded by $W(\Phi)$ is not clear,
and the dangerousness of the irrelevance of this operator remains mysterious to us.

\section{A family of examples of dilaton-driven confinement}
\label{app:solutions}

The solution in (\ref{gubsol}) is a particular member of a more general family of solutions of Type IIB supergravity. In this appendix we will present this family of solution.
The following is a saddle point of Type IIB SUGRA action (with $B^{NS}_{2} = C^{RR}_{2} = 0$) 
$$ ds^2 = L^{2}\(-\frac{{d\tau}^2 {\CK_x}}{r^2} + \frac{{d\vec{x}}^2 {\CK_x}}{r^2} +
 \frac{{dy }^2 {\CK_x} {{\J}}}{r^2} + \frac{{dr}^2}{r^2}\) + L^2ds^{2}_{S^{5}}$$ 
 \be F_{5} = 2L^{4}(1 + \star) \Omega_{5}, \quad \Phi = \Psi = {\mho \over 2} 
 \log \({ 1+r^4/r_0^4 \over {1-r^4/r_0^4}}\), 
 \quad \chi = 0   \label{typeIIB}\ee
when 
\be \CK_x  = \sqrt{1-r^8/r_0^8}~~~\text{    and   } ~~~
\J =  \frac{1+{r^4\over r_0^4}}{\(1-{r^4\over r_0^4}\)^{\sqrt{6-\mho^2}}}
~.\ee
That is, the above field configuration satisfies the following equations (we choose units with $8\pi G_{10} =1$):
$$
   \nabla_{M}\nabla^{M} \Phi = e^{2\phi} (\nabla_M \chi)^2  $$
$$   \nabla_{M}\(e^{2 \Phi}\nabla^{M}\)  \chi =0$$
   \be   {R}_{MN} = \frac{1}{2} \partial_M \Phi \partial_N \Phi + 
    \frac{1}{2} e^{2\phi} \partial_M \chi \partial_N \chi + 
    {1 \over 6} F_{M P_1 \ldots P_4} F_N{}^{P_1 \ldots P_4} 
 \label{TypeIIBEOM}\ee
 $$ F_{5} = \star F_{5}, \quad dF_{5} = 0$$
 $$ \int_{S^{5}} F_{5} = N T_{D3} = N \sqrt{\pi} \implies L^{8} = {N^{2} \over 4 \pi^{5}} $$
Let us first consider the case when $y$ is non compact; in this case, the solution preserves 3+1 dimensional Lorentz invariance in $t, \vec x, y$ when $\mho = \sqrt{6}$. All solutions preserve 2+1 dimensional Lorentz invariance. 
We studied 
the solution at $\mho  = \sqrt{6}$
previously \cite{MottKBJM};
it realizes Schr\"odinger symmetry asymptotically.


All solutions with $\mho \neq 0$ are singular at $r=\rs$. 
For any value of $\mho ~( 0 < \mho \le \sqrt{6})$, the metric is conformal to a regular metric.  Let us pick any solution from this family of solutions. The singularity in this solution can be resolved by ``uplifting'' the solution to a regular solution of 11 D supergravity. This can be done in more than one way. One way is to T-dualize along $y$ to get a solution of type IIA. This T-duality will modify the profile of the dilaton, but it does not remove the singularity. The singularity in the type IIA solution can be removed by oxidizing this type IIA solution to a regular solution of 11D supergravity. The dilaton field becomes the ``radion'' associated with the 11-dimensional circle \cite{Witten:1995ex}. 

An alternate way of ``uplifting'' the solution was used in \cite{MottKBJM} to resolve a similar singularity. Here, the $S^{5}$ part of the metric is written as a Hopf fiber over $\mathbb{CP}^{2}$. Then we can obtain a solution of type IIA supergravity by T-dualizing along the Hopf circle (say $\chi_{1}$). This solution of Type IIA supergravity can then be uplifted to 11D supergravity (see \cite{MottKBJM}). The uplifted solution is
 $$ds^2_{11} = e^{-\Psi/6} L^{2}\Big[  \(-\frac{{d\tau}^2 {\CK_x}}{r^2} + \frac{{d\vec{x}}^2 {\CK_x}}{r^2} +
 \frac{{dy }^2 {\CK_x} {{\J}}}{r^2} + \frac{{dr}^2}{r^2}\) + ds^2\(\mathbb{CP}^2\) +  d\chi_1^2 \Big]+ L^{2} e^{4\Psi/3} d\chi_2^2$$
\be F_{4} = 
 L^{4}\( {1\over 2} J \wedge J 
 + 2  J\wedge d\chi_{1}\wedge d\chi_{2} \)  \label{11Dsol}\ee
where $J$ is the K\"ahler form on $\mathbb{CP}^{2}$. 
The uplifted solution is regular. We can now get two solutions of type IIA from this 11 dimensional solution - (a) by reducing along  $\chi_1$ and (b) by reducing along $\chi_2$. The first
reduction produces a regular metric (with a smoothly shrinking circle) and a constant dilaton, while the second system has a metric with a curvature singularity and non-trivial dilaton profile. 
The second system is related to the type IIB solution in (\ref{typeIIB}) by T-duality.
The two type-II solutions are related by S-duality. 

\section{Supersymmetry analysis}
\label{app:susy}

In this section we analyze the supersymmetry of the background in (\ref{Lif0}) (a related analysis 
of Killing spinors 
appears in \cite{Cassani:2011sv}). We will assume $x^{3}$ to be non-compact in this section. In the following we will use use $M, M_{i}, N$ to denote spacetime (10D) indices and $a,b$ to denote vielbein indices. The conditions for a bosonic background (with $B^{NS}_{2} = C^{RR}_{2} = 0$) to preserve some supersymmetry are \cite{GauntlettSparks}:

\noindent
$ \mbox{Dilatino ($\lambda$) variation:}$ \be \quad \delta \lambda =  {i} \gamma^{N} {\CP}_{N} \epsilon  =0 \label{dilatino}\ee
$ \mbox{Gravitino ($\psi_{M}$) variation:}$ \be \quad \delta \psi_{M} = \left(\del_{M} + {1\over 4} \omega_{M}^{ab} \Gamma_{ab} - {i \over 2} {\CQ}_{M} \right)\epsilon + {i \over 192} \gamma^{M_{1}M_{2}M_{3}M_{4}} F_{M M_{1}M_{2}M_{3}M_{4}} \epsilon  = 0\ee
where we have combined the Majorana-Weyl fermions ($\epsilon^{1,2}$) of type IIB supergravity into a single complex Weyl spinor $\epsilon = \epsilon^{1} + i \epsilon^{2}$ (following \cite{GauntlettSparks}).   
The variables $\CQ$ and $\CP$ are defined as follows
 $$\CQ = -{1\over 2} e^{\Phi} d \chi, \quad \CP = {i\over 2} e^{{\Phi}}d\chi + {1\over 2} d\phi. $$   
Note that only $\CP_{3}$ and$\CQ_{3}$ are non-zero in our case. All other components of $\CQ$ and $\CP$ are zero\footnote{Supersymmetry of somewhat different null backgrounds with ${\CQ} = 0, {\CP} \neq 0$ were studied in \cite{NullCos}.}. Let us define $\tilde{Q} = QL e^{\Phi_{0}}$ for convenience.   $\Gamma$ denote ``flat space'' gamma matrices ({\ie} $\{\Gamma^{a}, \Gamma^{b} \} = 2 \eta^{ab}$) and $\gamma_{M} = e^{a}_{M} \Gamma_{a}$ are curved space gamma matrices. Note that only the non-compact part of the background in  (\ref{Lif0}) is different from the undeformed $AdS_{5} \times S^{5}$ background. Hence, we will suppress the $S^{5}$ part in the rest of the analysis. We have defined $e^{a}_{M}$ in section \ref{sec:nulldef}. The spin connections associated with this choice of orthonormal basis are
$$ \omega^{04} = 2 e^{0}, \quad \omega^{y4} = 2 e^{0}, \quad \omega^{i4} = e^{i}$$
Now, for the dilatino variation  \eqref{dilatino} to vanish, we must have $\gamma_{t} \epsilon  = 0 = \(\Gamma_{0} + \Gamma_{y}\) \epsilon$.  \footnote{Note that $\gamma_{t}^{2} =   \(\Gamma_{0} + \Gamma_{y}\)^{2} = 0$.}Using this constraint and the expressions for the spin connections, we can write the gravitino equations as follows 
$$ \del_{t} \epsilon = 0, \quad \(\del_{3} - {i \over 2} \CQ_{3} - {1 \over 2} e^{y}_{3}\Gamma_{y} \)\epsilon = 0,$$ $$\quad \(\del_{j} - {L \over 2 r} \Gamma_{j} \(1 - \Gamma_{4}\) \)\epsilon = 0, \del_{r} \epsilon - {L \over 2 r} \Gamma_{4} \epsilon =0 $$
Note that the last two equations are the same as the equations in undeformed $AdS_{5}$. Note that $\epsilon$ must satisfy $\(\Gamma_{0} + \Gamma_{y}\) \epsilon = 0$ and $\Gamma_{4} \epsilon = \epsilon$. The latter condition is the constraint we get from the last two equations (same as the undeformed case). We can see that the following satisfies all the equations.
$$ \epsilon = {e^{i\tilde{Q} x_{3} \over 4L_{3}} } \( \Gamma_{0} + \Gamma_{y} \)   \Bigg[ \({1 + \Gamma_{y}\over 2}\)  + e^{\tilde{Q} x_{3} \over 4L_{3}} \({1 - \Gamma_{y}\over 2}\) \Bigg] \( {1 + \Gamma_{4} \over 2}\) {L\over r^{1/2}}\eta  $$ 
where $\eta$ is independent of $t, x_{i}, x_{3}$ and $r$. When $\CQ = 0, \CP \neq 0$ this result agrees with the results of \cite{NullCos}. The above analysis shows that the background in (\ref{Lif0}) preserves 1/4 of the original supersymmetry. Hence, the operators dual to the null deformations in the bulk preserve 4 supercharges.

\section{Boundary terms}
\label{app:counterterms}

In this section, we will show that 
in the geometry \eqref{Lifcut},
$r_{0}$ 
(and hence the IR cutoff scale $r_{\star}$) 
is determined by non-trivial boundary conditions on $e^{y}$. We will do this by finding the boundary terms that are required to have a well-defined variational principle. As discussed in appendix 
\ref{sec:dilatondriven}, the parameter that determines the boundary behavior of $e^{y}$ corresponds to a mass-deformation in the dual field theory. Note that when $x^{3}$ is compact, the one point functions of the operators dual to $e^{y}_{3}$ and $e^{y}_{t}$ must be finite to have a well-defined variational principle in addition to finiteness of stress tensor and one point function of other supergravity fields such as the dilaton and axion. 
Finiteness of the five dimensional stress tensor (and other one point functions) in the non-compact case does not guarantee the finiteness of the one point functions in the case where $x_{3}$ is compact. In fact when $r_{0}$ is finite, the geometry ends in the IR; while in the non-compact case the geometry does not end in the interior and $r_{0}$ is just a parameter associated with the plane wave. We will be exploiting this crucial difference in the asymptotic behavior in this section to interpret the parameter $r_{0}$ as a mass deformation of the dual field theory (when $x_{3}$ is compact). 

When $x_{3}$ is compact, it is convenient to work with the reduced theory to find the boundary terms. The lower dimensional boundary terms need not oxidize to a local, intrinsic boundary term in the higher dimensional theory in general. It appears that in the case of gravity duals of dipole theories, the higher dimensional action contains boundary terms that are not local. In these cases,  it should be possible to obtain the ten-dimenional counterterms by performing a Melvin twist of the $AdS$ boundary terms. The action of the Melvin twist on the ten-dimensional Gibbons-Hawking term would produce unfamiliar extrinsic terms involving the $B_{NS}$, metric and the dilaton in addition to non-local intrinsic boundary terms. In fact application of holographic renormalization of  a ten-dimensional solution seems to be less understood when the internal manifold is squashed. 

In the present case, we will show that the lower dimensional boundary terms can be oxidized to local boundary terms in the higher dimensional theory when $r_{0} \rightarrow \infty$. When $r_{0}$ is finite, the lower dimensional counterterms do not seem to uplift to a local counterterm of the higher dimensional theory. Further, the boundary conditions on the higher dimensional metric are complicated due to the compactness of $x_{3}$.  It appears that a generalization of GH boundary term is required to include other non-trivial boundary behavior of the metric. Note that the Gibbons-Hawking boundary term imposes Dirichlet boundary condition on all the components of the metric. It is possible to impose other boundary conditions (without modifying the Gibbons-Hawking term) by introducing a heavy field (proxy for ``string field'') that induces the boundary condition on the metric components (as discussed in appendix \ref{sec:dilatondriven}). The fact that the higher dimensional counterterm needs to be modified when $r_{0}$ is finite is an indication that $r_{0}$ is determined by some boundary condition induced by an excited string state (or some heavy field).  

We must understand the boundary conditions on $e^{y}$. 
We will do this by fixing the boundary terms in the reduced theory. 
First we present the action of the reduced theory without the boundary terms: 
$$ S = {L_{3}\over 2 \kappa^{2}_{4}} \int d^{4}x \sqrt{{g}_{4}} \Bigg( {R}_{4}  -{1\over 2} \(\del \tilde{\Phi}\)^{2}- {3\over 2} \(\del \sigma\)^{2} -  \({Q^{2} \over2 L_{3}^{2}} e^{2\tilde{\Phi} - 3 \sigma} -{12 \over L^{2}} e^{-\sigma}\) $$ \be \qquad \qquad \qquad  \qquad \qquad \qquad  \qquad \qquad \qquad - {1\over 4}e^{3 \sigma} F^{2} - {Q^{2}  \over2 L_{3}^{2}}e^{2\tilde{\Phi}}A^{2} \Bigg) \label{redAct}\ee
where $\tilde{\Phi}$ is the scalar field obtained from dimensional reduction of the type IIB dilaton, $e^{2\sigma}$ is the radion field associated with $x_{3}$. Some details of the reduction can be found in \cite{LifString}.  The dimensional reduction of (\ref{Lifcut}) produces the following solution which is a saddle point of the reduced action \eqref{redAct}\footnote
{We note that this reduction was used in \cite{Amado:2011nd} to embed Lifshitz black holes 
in string theory.
}:
\be ds^{2} = e^{\sigma} L^2\( -e^{- 2\sigma } {dt^{2} \over r^{4}} + {d\vec{x}^{2} \over r^{2}} + {dr^{2} \over r^{2}} \), \quad A = {L^2 e^{-2\sigma} dt \over r^{2}}, \quad \tilde{\Phi} = \Phi_{0}, \quad e^{2\sigma} = f(r) \label{Lower}\ee
Note that the term quadratic in the vector field $A$ depends on $\tilde{\Phi}$. The linearized fluctuations of $A$ and $\sigma$ satisfy the equations in (\ref{fluct}). 

To determine the requisite boundary terms, we employ the following logic.
We demand that in the limit $r_\star \to\infty$, the boundary conditions
and boundary terms are the natural ones in 5d.  
Then we add intrinsic 4d counterterms which make the stress tensor finite 
with the same boundary conditions.
This will give physics consistent with the desired scheme in Fig.~\ref{fig:scheme}.

The 5d boundary terms (when $r_\star \to \infty$) are:
 \be S^{(5d)}_{bdy} = \int d^{4}x\sqrt{\gamma'} \(K'  - {3\over L}  - {e^{2\Phi} \over 4} \chi \Box_{\gamma'} \chi  \) \label{Highct} \ee
where $\Box_{\gamma'}$ is the Laplacian on the boundary metric $\gamma'$ (see \eg~\cite{Skenderis}). 
In particular, the 5d Gibbons-Hawking term $K'$ imposes Neumann boundary conditions on $\sigma$.

For the action in (\ref{redAct}) to be well-defined on (\ref{Lower}) we need to introduce the following boundary terms
$$ S_{bdy} = \int d^{3}x\sqrt{\gamma} \Bigg(K  + e^{-3 \sigma}n^{\mu} A^{\nu} F_{\mu \nu} + n^{r}\del_{r}\sigma W_{1}\({\sigma}, A^{2}\)  - {3\over L}e^{-\sigma/2} $$ \be \qquad \qquad  \qquad \qquad \qquad \qquad  \qquad \qquad+ W_{2}\({\sigma}, A^{2}, \Phi\) + W_{3}\({\sigma}, A^{2}, \Phi\) \Bigg) \label{lowbdy}\ee
where the functions $W_{1}$ , $W_{2}$ and $W_{3}$ are defined as follows
$$W_{1}\(\sigma, A^{2}\) = -{2\over L} e^{-\sigma/2} -{2\over L} e^{5\sigma/2} A^{2}$$
$$W_{2}\(\sigma, A^{2}, \Phi\) =  {Q^{2} L \over 8 L_{3}^{2}}e^{2 \Phi}\(e^{\sigma/2} A^{2} + e^{-5\sigma/2} \)  $$
$$W_{3}\(\sigma, A^{2}, \Phi\) = {1\over \kX^2} \Bigg(\({Q^{2} L^2 \over 4 L_{3}^{2}}e^{2 \Phi}e^{-2\sigma}-1\) \(L W_{1}\(\sigma, A^{2}\)+  \({Q^{3} L^{4} \over 8 L_{3}^{3}}e^{3 \Phi}\)A^{4}\)$$ $$\qquad \qquad \qquad \qquad \qquad \qquad+{1\over 2}\({Q^{2} L^2 \over 4 L_{3}^{2}}e^{2 \Phi}e^{-2\sigma}-1\)^{2} e^{-\sigma/2} \Bigg)~.$$
The first three terms of the lower dimensional boundary term come from the reduction of the higher dimensional Gibbons-Hawking term. The higher dimensional ``boundary cosmological constant'' reduces to the fourth term ($e^{-\sigma/2}$) in the lower dimensional boundary integral, and the last term of (\ref{Highct}) (the axion kinetic energy) reduces to $W_{2}$ term in the lower dimensional boundary action. 
This action without $W_3$ makes the stress tensor finite even when $r_{0}$ is finite.  
However, this does not make the variation with respect to $e^{y}_{3}$ and $e^{y}_{t}$ finite. 
This variation can be cancelled by adding $W_3$, with the specific coefficient
\be \kX =r_{\star}/4 = r^{2}_{0}e^{\Phi_0}Q /2L_{3} .\ee 
Note that $W_3$ cannot be lifted to an intrinsic local 5d counterterm.
We interpret $W_{3}$ as a boundary term that is induced by a ``string field'' which is not directly visible in supergravity. 
Further, when $\kX$ or $r_{0}$ is infinite, the lower dimensional boundary terms can be uplifted to the 5D boundary term in 
(\ref{Highct}). 

{\cog{Let us make few comments about the interpretation of $r_0$ when $x_3$ is non-compact, just as an aside.
The 5D boundary term in (\ref{Highct}) makes the five dimensional stress tensor finite even when $r_{0}$ is finite. We would like to emphasize again that this does not make the variation with respect to $e^{y}_{3}$ and $e^{y}_{t}$ finite. This 5D boundary term does not contain any term associated with $r_0$. Hence, $r_0$ is a parameter specifying the state, when $x_3$ is non-compact. However, this interpretation is correct only when the boundary terms do not depend on $r_0$. }}

We emphasize the distinction between $\Gamma \sim \kappa_6 \sim \kappa_8$, which
determines the coefficient in the perturbed action of operators
whose dimensions are protected by supersymmetry above the KK scale (see Fig.~\ref{fig:scheme}), 
and $M_{\text{BC}}$ which cannot be interpreted in this way and 
sources a ``string field".

 \end{document}